%% file: new-vldb.tex
\documentclass[sigconf, nonacm]{acmart}

\newcommand\vldbdoi{XX.XX/XXX.XX}
\newcommand\vldbpages{XXX-XXX}
\newcommand\vldbvolume{18}
\newcommand\vldbissue{1}
\newcommand\vldbyear{2025}

\newcommand\vldbtitle{\shorttitle} 
\newcommand\vldbavailabilityurl{https://github.com/xdyangsh/FastKCNA}
\newcommand\vldbpagestyle{plain} 

\usepackage{weiwAlgorithm}
\usepackage{makecell}

\DeclareMathOperator*{\argmin}{arg\,min}

\newcommand{\eat}[1]{}
\newcommand\blfootnote[1]{%
  \begingroup
  \renewcommand\thefootnote{}\footnote{#1}%
  \addtocounter{footnote}{-1}%
  \endgroup
}

\input{custom_sytles}

\begin{document}
\title{Revisiting the Index Construction of Proximity Graph-Based Approximate Nearest Neighbor Search}

\author{Shuo Yang$^\S$}
\affiliation{%
  \institution{Xidian University}
  \country{}
}
\email{yangsh@stu.xidian.edu.cn}

\author{Jiadong Xie$^\S$}
\affiliation{%
  \institution{The Chinese University of Hong Kong}
  \country{}
}
\email{jdxie@se.cuhk.edu.hk}

\author{Yingfan Liu$^*$}
\affiliation{%
  \institution{Xidian University}
  \country{}
}
\email{liuyingfan@xidian.edu.cn}

\author{Jeffrey Xu Yu}
\affiliation{%
  \institution{The Chinese University of Hong Kong}
  \country{}
}
\email{yu@se.cuhk.edu.hk}

\author{Xiyue Gao}
\affiliation{%
  \institution{Xidian University}
  \country{}
}
\email{xygao@xidian.edu.cn}

\author{Qianru Wang}
\affiliation{%
  \institution{Xidian University}
  \country{}
}
\email{wangqianru@xidian.edu.cn}

\author{Yanguo Peng}
\affiliation{%
  \institution{Xidian University}
  \country{}
}
\email{ygpeng@xidian.edu.cn}

\author{Jiangtao Cui}
\affiliation{%
  \institution{Xidian University}
  \country{}
}
\email{cuijt@xidian.edu.cn}

\begin{abstract}

Proximity graphs (PG) have gained increasing popularity as the state-of-the-art solutions to $k$-approximate nearest neighbor ($k$-ANN) search on high-dimensional data, which serves as a fundamental function in various fields, e.g., retrieval-augmented generation. Although PG-based approaches have the best $k$-ANN search performance, their index construction cost is superlinear to the number of points. Such superlinear cost substantially limits their scalability in the era of big data. Hence, the goal of this paper is to accelerate the construction of PG-based methods without compromising their $k$-ANN search performance.

To achieve this goal, two mainstream categories of PG are revisited: relative neighborhood graph (RNG) and navigable small world graph (NSWG). By revisiting their construction process, we find the issues of construction efficiency. To address these issues, we propose a new construction framework with a novel pruning strategy for edge selection, which accelerates RNG construction while keeping its $k$-ANN search performance. Then, we integrate this framework into NSWG construction to enhance both the construction efficiency and $k$-ANN search performance of NSWG. Extensive experiments are conducted to validate our construction framework for both RNG and NSWG, and that it significantly reduces the PG construction cost, achieving up to 5.6x speedup, while not compromising the $k$-ANN search performance.
 
\end{abstract}

\maketitle

\blfootnote{$^\S$ Shuo Yang and Jiadong Xie are the joint first authors.}
\blfootnote{$^*$ Yingfan Liu is the corresponding author.}

\pagestyle{\vldbpagestyle}
\begingroup\small\noindent\raggedright\textbf{PVLDB Reference Format:}\\
Shuo Yang, Jiadong Xie, Yingfan Liu, Jeffrey Xu Yu, Xiyue Gao, Qianru Wang, Yanguo Peng, and Jiangtao Cui. \vldbtitle.
PVLDB, \vldbvolume(\vldbissue): \vldbpages, \vldbyear.\\
\href{https://doi.org/\vldbdoi}{doi:\vldbdoi}
\endgroup
\begingroup
\renewcommand\thefootnote{}\footnote{\noindent
This work is licensed under the Creative Commons BY-NC-ND 4.0 International License. Visit \url{https://creativecommons.org/licenses/by-nc-nd/4.0/} to view a copy of this license. For any use beyond those covered by this license, obtain permission by emailing \href{mailto:info@vldb.org}{info@vldb.org}. Copyright is held by the owner/author(s). Publication rights licensed to the VLDB Endowment. \\
\raggedright Proceedings of the VLDB Endowment, Vol. \vldbvolume, No. \vldbissue\ %
ISSN 2150-8097. \\
\href{https://doi.org/\vldbdoi}{doi:\vldbdoi} \\
}\addtocounter{footnote}{-1}\endgroup

\ifdefempty{\vldbavailabilityurl}{}{
\vspace{.3cm}
\begingroup\small\noindent\raggedright\textbf{PVLDB Artifact Availability:}\\
The source code, data, and/or other artifacts have been made available at \url{\vldbavailabilityurl}.
\endgroup
}

\section{Introduction}

Recent breakthroughs in deep learning models, specifically embedding models, have revolutionized the representation of various types of data, such as images and text chunks. These models transform the data into vectors, which encapsulate key information in a high-dimensional space for semantic analysis. As a result, $k$-approximate nearest neighbor ($k$-ANN) search on high-dimensional vectors has emerged as a fundamental problem across multiple domains, including information retrieval~\cite{GrbovicC18,HuangSSXZPPOY20}, recommendation system~\cite{OkuraTOT17} and large language models~\cite{barthel2019real,RAG,REALM,RAG-ACL,liu2024retrievalattention}. In particular, due to the popularity of retrieval-augmented generation~(RAG) \cite{RAG, RAG-ACL}, $k$-ANN search on high-dimensional vectors causes more and more attention as a key RAG component. 
Given a dataset $D \subset \mathbb{R}^d$ and a query vector $q \in \mathbb{R}^d$, $k$-ANN search returns the $k$ vectors in $D$ that are sufficiently close to $q$, where $k$ is a specified parameter.

$k$-ANN search has been extensively studied for several decades, hence there exists a wealth of methods in the literature~\cite{Dpg,survey2021, Wang2017PAMI,Sw,Overlap, LSB, C2LSH, SKLSH}.
According to recent studies~\cite{AumullerBF20,LiZAH20,Dpg,survey2021}, the proximity graph (PG) based methods
\cite{Overlap,KGraph,Sw,Hnsw,Dpg,Nsg,Nssg,taumg}
have emerged as the promising solution and have demonstrated superior performance compared to other approaches, such as hashing-based methods \cite{MPlsh, LSB, C2LSH, SKLSH, SRS, QALSH}, invert index-based methods \cite{PQ, IMI} and tree-based methods \cite{Kdtree, Xtree, Srtree, Flann}.
A PG treats each vector $u \in D$ as a graph vertex and then connects edges between $u$ and its close neighbors. Notably, all PGs share the same vertex set but have various edge sets due to their distinct edge-selection strategies.
According to previous studies~\cite{taumg, survey2021, Dpg,Pan0L24}, PGs are divided into three categories by different neighbor-selection strategies: $k$ nearest neighbor graph (KNNG), relative neighborhood graph (RNG), and navigable small world graph (NSWG).

Although PG-based approaches have superior search performance, they still suffer from a significantly higher cost in terms of index construction than other methods. 
This is primarily due to the necessity of identifying close neighbors for each point to establish the graph edges. 
{In the traditional applications of $k$-ANN search, PG index is built offline and then responds to $k$-ANN queries. Thus its construction cost is treated as a second-class performance indicator. However, in the emerging scenario, i.e., the RAG model training~\cite{REALM, ATLAS, RAG-ACL}, the PG index will be frequently built in an online manner, due to the tuning on the embedding model that transforms the source data such as text chunks into vectors. As a result, it is urgent to build PG efficiently while maintaining search performance.}
To address this issue, several studies have been proposed to accelerate the index construction, such as DiskANN~\cite{diskann}, RNN-Descent~\cite{OnoM23}, LSH-APG~\cite{Lsh-apg} and ParlayANN~\cite{ManoharSBD0S024}. However,
as shown in Table~\ref{tab:compare},
their search performance remains noticeably inferior to NSG (Navigating Spread-out Graph)~\cite{Nsg}, where NSG is one of the state-of-the-art (SOTA) methods that do not prioritize index construction.
In Table~\ref{tab:compare} on \kw{Gist1M}, the second column shows the time of index construction and the next four columns depict queries per second (QPS) at different recall levels (higher is better). Hence, they sacrifice the search performance to expedite the index construction, which violates the primary goal of $k$-ANN search.

\begin{table}[t]
    \centering
    \caption{PG index comparisons on Gist1M}
    \vspace{-3mm}
    \label{tab:compare}
    \resizebox{0.85\linewidth}{!}{
    \begin{tabular}{|c|c|c|c|c|c|}
        \hline
        \multirow{2}{*}{ Approaches }   & \multirow{2}{*}{\makecell[c]{Building\\Time (s)}}&\multicolumn{4}{c|}{QPS when Recall@10=} \\ \cline{3-6}
        &&0.85&0.90&0.95&0.99 \\ \hline \hline
        NSG~\cite{Nsg} &573&\textbf{971}&\textbf{712}&\textbf{417}&{133} \\ \hline
        LSH-APG~\cite{Lsh-apg} &433&453&305&172&53 \\ \hline
        DiskANN~\cite{diskann} &214&657&469&266&83 \\ \hline
        ParlayANN~\cite{ManoharSBD0S024} &409&669&472&274&89 \\ \hline
        RNN-Descent~\cite{OnoM23} &\textbf{121}& 739&539&345&\textbf{139}\\ \hline
        \rowcolor[gray]{0.95}
        FastNSG (ours) & {119}&{964}&{694}&{402}&{148} \\ \hline
    \end{tabular}}
    \vspace{-5mm}
\end{table}


In this work, we begin by revisiting the construction process of SOTA PGs, namely RNG and NSWG. 
On top of the KNNG, RNG is built with three key phases. 
In the first phase, an initial KNNG $G_{k_0}$ is constructed (\kw{initialization} phase), and then the results of $k$-ANN search on $G_{k_0}$ for each node are obtained as candidate neighbors (\kw{search} phase), which are further refined to prune redundant edges and enhance connectivity (\kw{refinement} phase). Unlike RNG assuming the priori knowledge of the whole dataset, NSWG builds from scratch and inserts nodes incrementally into the current graph one by one. Like RNG, NSWG shares both \kw{search} phase that finds close neighbors in the current incomplete graph index and \kw{refinement} phase that builds edges between each node and a selected subset of those neighbors found. 

Up on the revisiting, we identify the PG construction issues and further propose an efficient construction framework for RNG, which could also be used to accelerate the NSWG index construction. To be specific, our contributions are summarized as follows.
\ding{202} 
During revisiting the index construction procedures of current PGs,
we analyze the essential operation in PG construction, i.e., {size-{$k$} {c}andidate {n}eighbor set {a}cquisition} ($k$-CNA), whose quality is key to the finally built PG.
RNG obtains $k$-CNA results by \kw{initialization} phase and \kw{search} phase, while NSWG by \kw{search} phase.
However, we find issues related to $k$-CNA, i.e., (1) the inefficiency of obtaining $k$-CNA results in RNG, and (2) the poor $k$-CNA quality in NSWG. 
\ding{203} To tackle the inefficiencies of $k$-CNA in RNG, we introduce a new RNG construction framework featuring a novel pruning strategy aimed at improving the efficiency of $k$-CNA while preserving its quality.
\ding{204}
To enhance the quality of $k$-CNA in NSWG, primarily caused by its node-by-node insertion method, we substitute this method with a layer-by-layer insertion strategy, and integrate it with our RNG construction framework to further enhance its efficiency.
\ding{205} We conduct extensive experiments on real-life datasets to validate the effectiveness of our construction framework. The results demonstrate that our framework accelerates the construction of the representative RNG index NSG and the representative NSWG index HNSW (Hierarchical Navigable Small World) up to 5.6x and 4.6x respectively, while achieving comparable or even better search performance.

The paper is organized as follows. We provide the preliminaries in Section~\ref{sec:pre}. In Section~\ref{sec:rng}, we revisit the existing PG methods and identify their construction issues. In Section \ref{sec:new-strategy}, we propose the refinement-before-search scheme as the basis of our construction framework for RNG and NSWG as shown in Section \ref{sec:new-framework}. 
We present our experimental studies in Section~\ref{sec:exp}.
Furthermore, we discuss the related works in Section~\ref{sec:rew} and conclude our work in Section~\ref{sec:clu}.

\section{$k$-ANN Search and Proximity Graphs}
\label{sec:pre}

\stitle{$k$-ANN Search:}
Let $D \subset \mathbb{R}^d$ be a high-dimensional dataset consisting of $n$ $d$-dimensional points. We denote the L2 norm (i.e., Euclidean distance) between two points $u, v \in \mathbb{R}^d$ as $dist(u,v)$.
Given a dataset $D$ and a query point $q\in \mathbb{R}^d$, \textit{the $k$-approximate nearest neighbor ($k$-ANN)} search aims to find the top-$k$ points in $D$ with the minimum distances from the query $q$. This paper focuses on the in-memory solutions, which assume that $D$ and the corresponding index can be hosted in the memory~\cite{Sw,Hnsw,Nsg,Dpg,Nssg,taumg}.
Table~\ref{tab:notation} summarizes the notations.

According to recent studies~\cite{AumullerBF20, LiZAH20, Dpg, survey2021}, proximity graph (PG) based approaches are the SOTA methods for $k$-ANN search. 
In the following, we discuss PG-based approaches and their search algorithm for answering $k$-ANN queries.

\begin{table}
\centering
\caption{Summary of notations}
\vspace{-3mm}
\label{tab:notation}
\resizebox{0.9\linewidth}{!}
{
\begin{tabular}{|r|l|}
\hline
 & {Definition} \\
\hline \hline
$D$ & the set of $n$ $d$-dimensional vectors \\
\hline
$G = (V, E)$ & a proximity graph with vertex set $V$ and edge set $E$ \\
\hline
$N_G(u)$ & the set of out-neighbors of $u$ in $G$ \\
\hline
$V(G)$/$E(G)$ & the node/edge set of $G$ \\
\hline
$q$ & a query data point\\
\hline
$k$ & the number of returned results in $k$-ANN search \\
\hline 
$L$ & the pool width in $k$-ANN search of PG \\
\hline
$ep$ & the entry point in $k$-ANN search of PG \\
\hline
$C(u)$ & the $k$-CNA results of $u$ \\
\hline
$M$ & the upper bound of node out-degrees in PG \\
\hline
$ef$ & the key construction parameter of HNSW \\
\hline
$\alpha$ & the angle threshold in $\alpha$-pruning \\
\hline
\end{tabular}}
\vspace{-5mm}
\end{table}
\subsection{Proximity Graph}
A PG $G=(V,E)$ of $D$ is a directed graph, where each node in $V$ uniquely represents a vector (i.e., data point) in $D$, and two nodes are connected by an edge in $E$ if their corresponding vectors are close to each other.
For a node $u \in V$, we use $N_G(u)$ to denote the set of out-neighbors of node $u$ in a PG $G$.

As presented in previous studies~\cite{taumg, survey2021, Dpg,Pan0L24}, PGs are classified into three categories according to their edge-selection strategies.

\vspace{1mm}
\stitle{$k$-Nearest Neighbor Graph (KNNG):}
Each node in KNNG is connected to its $k$-approximate nearest data points, which is proposed as an approximation of the Delaunay graph (DG)~\cite{dg}, since DG becomes a complete graph when the dimension is large~\cite{fanng,taumg}.

\vspace{1mm}
\stitle{Relative Neighborhood Graph (RNG):}
RNG is constructed based on KNNG, it removes the longest edge in every possible triangle in the KNNG, i.e., if edge $(u,v)$ exists in RNG, there exists no edge $(u,w)$ in the graph such that $dist(u,w)<dist(u,v)$ and $dist(v,w)<dist(u,v)$.
RNG enhances its connectivity by adding extra edges between connected components, while limiting the out-degree of each node to a small constant~\cite{Nsg}.

\vspace{1mm}
\stitle{Navigable Small World Graph (NSWG):}
NSWG is derived from Milgram's social experiment~\cite{travers1977experimental}, which demonstrates that two nodes in a large graph are connected by a short path that can be discovered through greedy routing. Hence, the NSWG construction involves incrementally node-by-node insertion, where each node is connected to its $k$-ANN in the current incomplete graph.

\subsection{Beam Search}

Although PGs have different construction algorithms, they share the same search method called beam search, which is on top of greedy routing. 
Greedy routing involves selecting the nearest out-neighbor of the current node at each step and terminates when no closer out-neighbors are available. 


To extend the greedy routing for $k$-ANN search, beam search (i.e., the best-first search) is thus proposed.
Specifically, as shown in Algorithm~\ref{alg:knn_search}, the search process starts from an entering point $ep$ and puts it in a sorted array $pool$ of nodes, which is maintained to store the currently found $L$-closest neighbors (Lines 1-2). Then, it iteratively extracts the closest but unexpanded neighbor $u$ from $pool$ (Line 4) and expands $u$ to refine $pool$, until the termination condition is satisfied (Line 3).
In each iteration, expanding $u$ for $q$ is shown in Lines 5-7, where each neighbor $v \in N_G(u)$ is treated as a $k$-ANN candidate of $q$ (Line 5) and further verified by an expensive distance computation  (Line 6) to refine $pool$ (Line 7). At the end of each iteration (Line 8), the algorithm finds the closest but unexpanded vertex in $pool$ as the next one to be expanded. It terminates when the first $L$ vertices in $pool$ have been expanded (Line 3).

\begin{algorithm}[t]
\small  
    \SetVline 
    \SetFuncSty{textsf}
    \SetArgSty{textsf}
 \caption{\texttt{KANNSearch}($G, q, k, L, ep$)}
 \label{alg:knn_search}
 \Input{graph index $G$, query point $q$, $k$ for top-$k$, pool width $L$ and entering point $ep$}
 \Output{$k$-ANN of query point $q$}
\State{$i \leftarrow 0$}
\State{$pool[0] \leftarrow (ep,dist(q, ep))$}
\While{$i < L$} 
{
	\State{$u\leftarrow pool[i]$}
	\For{each {$v \in N_G(u)$}}
        {
		\State{insert $(v, dist(q, v))$ into $pool$}
	}
	\State{sort $pool$ and keep the $L$ closest neighbors}
	\State{$i \leftarrow $ index of the first unexpanded vertex in $pool$}
}
\Return{$pool[0, \ldots, k-1]$}
\end{algorithm}

We can see that $L$ plays a crucial role in the trade-off between efficiency and accuracy, i.e., a higher $L$ leads to more accurate results but a larger time cost.
Furthermore, beam search is a key component applied to answer $k$-ANN search in the PG construction, as discussed in the next section. 

\section{PG Construction Revisited}
\label{sec:rng}

In this section, we revisit the index construction of the SOTA PG approaches, namely RNG methods~\cite{Nsg,Nssg,taumg,diskann} and NSWG methods~\cite{Hnsw,Sw,Lsh-apg}.
We first present the construction process of RNG and NSWG respectively, followed by identifying their issues in building efficiency. 

\subsection{RNG Index Construction}
To construct an RNG index, the computation of distances between each pair of nodes is necessary in order to eliminate the longest edge in all possible triangles. Unfortunately, its time complexity is $O(n^2d)$, which is practically impossible for large data. Hence, some studies propose to construct a practical version of RNG via another PG, such as KNNG~\cite{Nsg, Nssg}, NSG~\cite{taumg} or even a random graph~\cite{diskann}.

In this section, we take NSG as an RNG representative. This is because NSG is widely acknowledged as one of the SOTA methods for $k$-ANN search~\cite{taumg, survey2021}, and recent studies focus on improving search performance by making minor modifications on NSG, e.g., extending the pruning strategy~\cite{Nssg, taumg}.

\begin{figure}
    \centering
    \subfigure[RNG-based approaches]
    {
    \includegraphics[width=\linewidth]{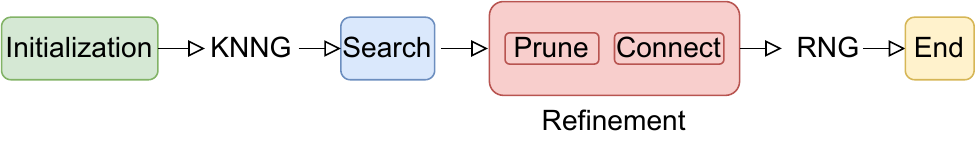}
    \label{fig:build_nsg}
    }
    \subfigure[NSWG-based approaches]
    {
\includegraphics[width=0.88\linewidth]{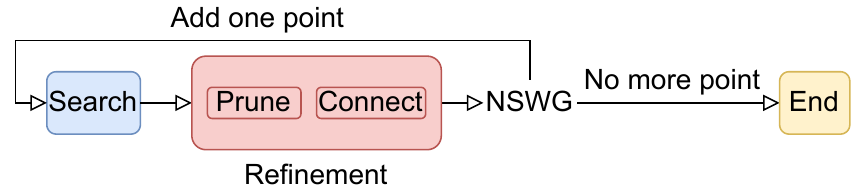}
\label{fig:build_hnsw}
    }
    \subfigure[Our framework]
    {
    \includegraphics[width=0.95\linewidth]{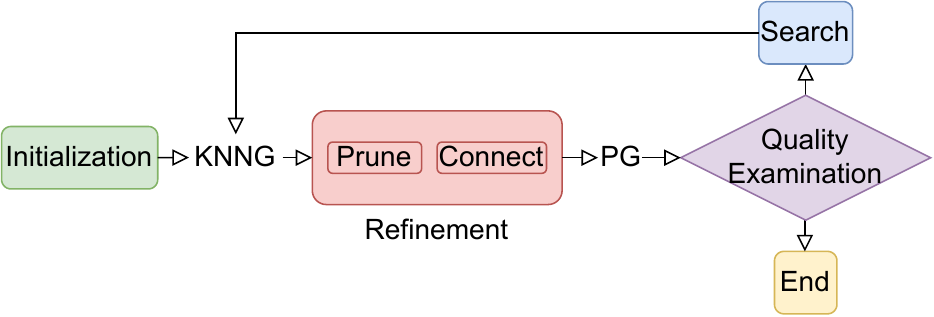}
\label{fig:build_pg}
    }
    \vspace{-3mm}
    \caption{The construction phases}
    \vspace{-3mm}
    \label{fig:build-phases}
\end{figure}




In a nutshell, as shown in Figure~\ref{fig:build_nsg}, the construction of NSG involves three phases: \kw{initialization}, \kw{search} and \kw{refinement}. The \kw{initialization} phase focuses on constructing an approximate KNNG, while \kw{search} phase aims to improve the quality of the KNNG, i.e., enhancing the accuracy of neighbors ($k$-ANN) for each node in the graph. Finally, the \kw{refinement} phase incorporates \kw{prune} and \kw{connect} operations to reduce the node out-degree and enhance the graph connectivity respectively.
We present the details of NSG construction process in Algorithm \ref{alg:build_nsg}.

\vspace{1mm}
\stitle{Initialization Phase:}
An initial KNNG $G_{k_0}$ where each node has $k_0$ neighbors is built in \kw{initialization} phase by the SOTA method called KGraph \cite{KGraph}, as shown in Line 1 of Algorithm \ref{alg:build_nsg}. The purpose of this phase is to build an index for \kw{search} phase. 

\eat{
Due to the time-consuming nature of constructing the exact KNNG, which requires $O(n^2d)$ time, several works~\cite{KGraph, FuC16,WangWZTGL12} efficiently builds approximate KNNG.
These algorithms all consist of two steps: \emph{initialize} and \emph{refine}.
In the \emph{initialize} step, KGraph~\cite{KGraph} randomly selects the initial $k$ neighbors, while \cite{WangWZTGL12} and EFANNA~\cite{FuC16} utilizes a set of tree structures to obtain the initial $k$ neighbors with higher quality.
In the \emph{refine} step, these algorithms take the neighborhood propagation strategies, which iteratively refine the $k$ neighbors by examining the second-order out-neighbors of each node. 

Among those methods, KGraph presents the best performance in KNNG construction, as shown in a recent experimental study \cite{KnngcSurvey}. Hence, as shown in Algorithm~\ref{alg:build_nsg}, the {Initialization} phase of NSG utilizes KGraph~\cite{KGraph} to create an initial approximate KNNG $G_{k_0}$ with $k_0$ neighbors for each $u \in D$ (Line 1).
}


\vspace{1mm}
\stitle{Search Phase:}
As in Lines 3-4 of Algorithm~\ref{alg:knn_search}, for each $u \in D$, the \kw{search} phase performs $k$-ANN search on $G_{k_0}$ in order to obtain the candidate set $C(u)$ for \kw{refinement}. Notably, each $k$-ANN search starts from the entering point $ep$ which is the closest point in $D$ to the centroid of $D$ (Line 2).

\vspace{1mm}
\stitle{Refinement Phase:}
This phase removes redundant neighbors in the set $C(u)$ via a pruning strategy (Algorithm \ref{alg:prune}) to obtain $N_G(u)$ with the constraint $|N_G(u)| \leq M$, where $M$ is a specific threshold (Lines 5-6). To improve the graph connectivity, unidirectional edges are added between $u$ and each $v \in N_G(u)$, which might trigger an extra pruning process in order to limit the out-degree of $u$ (Lines 7-8). 
Finally, a depth-first search (DFS) is employed to identify any remaining connected components in $G$, and additional edges are then added to connect them together (Lines 10-11).


\begin{algorithm}[t]
\small  
    \SetVline 
    \SetFuncSty{textsf}
    \SetArgSty{textsf}
\caption{\texttt{BuildNSG}($D,k_0, k, L, M$)}
\label{alg:build_nsg}
\Input{dataset $D$ and four parameters $k_0$, $k$, $L$ and $M$}
\Output{an NSG $G$}
\tcc{\textbf{Phase 1:} \texttt{Initialization}}
\State{build $G_{k_0}$ with $k_0$ neighbors by KGraph \cite{KGraph}}
\tcc{\textbf{Phase 2:} \texttt{Search}}
\State{$ep\leftarrow$ \texttt{KANNSearch} ($G_{k_0}, cn, k, L, rn$), where $cn$ is the centroid of $D$ and $rn \in D$ is a random node}
\For{{each $u \in D$ \underline{in parallel}}}
{
    \State{$C(u)\leftarrow$ \texttt{KANNSearch} ($G_{k_0}, u, k, L, ep$)}
}
\tcc{\textbf{Phase 3:} \texttt{Refinement} ($G =$\kw{Refine}$(\{C(u)|u\in D\}, M)$)}
\For{{each $u \in D$ \underline{in parallel}}}
{
{\State{$N_G(u)\leftarrow$ \kw{Prune}($u,C(u),M$)}}
}
\For{{each $u \in D$ \underline{in parallel}}}
{
    \State{$N_G(u)\leftarrow$ \kw{Prune}($u,N_G(u)\cup \{v|u\in N_G(v)\},M$)}
}
\State{find connected components via DFS}
\State{add extra edges into $E(G)$ between connected components}
\Return{$G$}
\end{algorithm}

\begin{algorithm}[t]
\small  
    \SetVline 
    \SetFuncSty{textsf}
    \SetArgSty{textsf}
\caption{\texttt{Prune}($u,C(u),M$)}
\label{alg:prune}
\Input{a vertex $u$, neighbor set $C(u)$ and out-degree limit $M$}
\Output{a pruned neighbor set of $u$}
\State{$PrunedNeighbor\leftarrow \varnothing$}
\For{each $v \in C(u)$ in the ascending order of $dist(u,v)$}
{
    \State{$DominateFlag\leftarrow$ false}
    \For{each $w \in PrunedNeighbor$}
    {
        \If{$dist(v,w)<dist(u,v)$}
        {
            \State{$DominateFlag\leftarrow$ true}
        }
    }
    \If{$DominateFlag=$ false}
    {$PrunedNeighbor\leftarrow PrunedNeighbor\cup \{v\}$}
    \State{\textbf{if} $|PrunedNeighbor|\ge M$ \textbf{then break}}
}
\Return{$PrunedNeighbor$}
\end{algorithm}

The widely used pruning process focuses on eliminating the longest edge within each possible triangle formed by the points in the dataset. 
For simplicity, we call this strategy as \emph{RNG pruning}. 
Specifically, if edge $(u,v)$ exists in the NSG only if $v$ is not dominated by any neighbor $w$ of $u$, i.e., there is no edge $(u,w)$ such that $dist(u,w)<dist(u,v)$ and $dist(v,w)<dist(u,v)$. In the practical version of NSG, this pruning process for each node $u$ is modified in two aspects, i.e., (1) the out-neighbors of each $u$ are only picked from the close neighbor set $C(u)$ and (2) each node has at most $M$ out-neighbors. The first modification improves the construction efficiency, while the second accelerates $k$-ANN search in NSG. 
The details of are presented in Algorithm~\ref{alg:prune}. Each candidate neighbor $v \in C(u)$ is checked individually, in ascending order of $dist(u,v)$ (Line 2). $v$ is selected as a neighbor of $u$ only if it is not dominated by any existing neighbors (Lines 3-8), and the process terminates once $M$ neighbors have been selected (Line 9). 

As to other RNG methods such as DPG \cite{Dpg} and $\tau$-MNG \cite{taumg}, we can build them by only replacing the RNG pruning strategy (Algorithm \ref{alg:prune}) with their own ones. Hence, the mainstream RNG methods follow the same framework of index construction.

\subsection{NSWG Index Construction}
\label{ssec:build_nswg}

Unlike RNG, NSWG is constructed by incrementally inserting nodes into the current graph and connecting each node to a subset of its $k$-ANN found in the current graph. Following this idea, several methods such as NSW~\cite{Sw}, HNSW~\cite{Hnsw} and LSH-APG~\cite{Lsh-apg} are proposed, where HNSW~\cite{Hnsw} stands out as the SOTA approach.
Hence, in this section, we focus on the HNSW construction process.

\begin{algorithm}[t]
\small  
    \SetVline 
    \SetFuncSty{textsf}
    \SetArgSty{textsf}
\caption{\texttt{BuildHNSW}($D,ef,M$)}
\label{alg:build_hnsw}
\Input{dataset $D$ and two parameters $ef$ and $M$}
\Output{an HNSW $G$}
\State{$m_L \leftarrow 0$}
\State{initialize $G_0$ with a randomly selected point $v\in D$ and no edges}
\For{{each $u\in D\setminus \{v\}$ \underline{in parallel}}}
{
    \State{randomly determine the highest layer of $u$ is $l$}
    \If{$l > m_L$}
    {
        \State{$m_L \leftarrow l$; $ep \leftarrow u$}
    }
    \tcc{\textbf{Phase 1:} \texttt{Search}}
    \State{$w\leftarrow ep$}
    \For{each $i\leftarrow m_L$ downto $l+1$}
    {
    \State{$w\leftarrow$ \texttt{KANNSearch}($G_i,u,1,1,w$)}
    }
    \State{$W_{l+1} \leftarrow \{w\}$}
    \For{$i\leftarrow l$ downto $0$}
    {
        \State{$W_{i}\leftarrow$ \texttt{KANNSearch}($G_i,u,ef,ef,W_{i+1}[0]$)}
    }
    \tcc{\textbf{Phase 2:} \texttt{Refinement}}
    \For{$i\leftarrow l$ downto $0$}
    {
        \For{each $u\in V(G_i)$}
        {
        \State{$N_{G_i}(u)\leftarrow$\kw{Prune}($u,W_i,M$)}
        }
        \For{each $u\in V(G_i)$}
        {
        \State{$N_{G_i}(u)\leftarrow$\kw{Prune}($u,N_{G_i}(u)\cup \{v|u\in N_{G_i}(v)\},M$)}}
    }
}
\Return{$G = \{G_0, G_1, \ldots, G_{m_L}\}$}
\end{algorithm}

As illustrated in Figure~\ref{fig:build_hnsw}, the construction of HNSW involves two phases for each inserted point: \kw{search} and \kw{refinement}. The details are presented in Algorithm~\ref{alg:build_hnsw}.
HNSW begins by initializing the graph with a single point (Lines 1-2). For each remaining point, HNSW randomly determines its highest layer $l$ using an exponentially decaying probability distribution (Line 4). Like NSG, the \kw{search} phase focuses on finding a candidate neighbor set for each node $u$. It starts the search from the top layer down to layer $l+1$ via greedy routing (Lines 8-9) and performs $k$-ANN search on each lower layer to obtain the candidate neighbor set for \kw{refinement} (Lines 10-12).
Next, for each lower layer (from $l$ to $0$), HNSW applies the RNG pruning strategy (Algorithm \ref{alg:prune}) to prune the neighbors obtained from the \kw{search} phase. Like NSG, HNSW adds undirectional edges between the inserted node and its selected neighbors to enhance connectivity, while limiting the out-degree of each node to a specific number $M$ (Lines 16-17). However, there is no \kw{connect} operation in the \kw{refinement} phase of HNSW.

\subsection{PG Construction Issues}
\label{sec:issues}

As discussed above, we can see that both RNG and NSWG construction share the same procedure for each node $u \in D$, i.e., finding a set of $k$ close neighbors of $u$, denoted \textit{size-\underline{$k$} \underline{c}andidate \underline{n}eighbor set \underline{a}cquisition} ($k$-CNA), {and then derive the PG by the \kw{refinement} phase that takes the $k$-CNA results as input. Specifically, the RNG construction method combines \kw{initialization} phase and \kw{search} phase to generate the $k$-CNA results, while the \kw{search} phase of NSWG construction performs $k$-ANN search on the current incomplete graph index for $k$-CNA results. Hence, \emph{both RNG and NSWG are derived from the $k$-CNA results in \kw{refinement} phase.}

Due to the importance of the $k$-CNA results, there are two performance aspects of obtaining $k$-CNA, i.e., efficiency and quality. First, the cost of obtaining $k$-CNA results contributes to the total construction cost and thus its efficiency is key to the construction efficiency. Second, the $k$-CNA quality, measured by the average over the recall of the $k$-CNA neighbors for each node $u \in D$ w.r.t. its exact $k$ nearest neighbors, significantly affects the $k$-ANN search performance of the graph index derived from $k$-CNA results. 
We present such an effect through the following experimental study.
}

\begin{figure}
  \centering
    \subfigure[NSG on \kw{Sift1M}]
    {
        \includegraphics[width=0.47\linewidth]{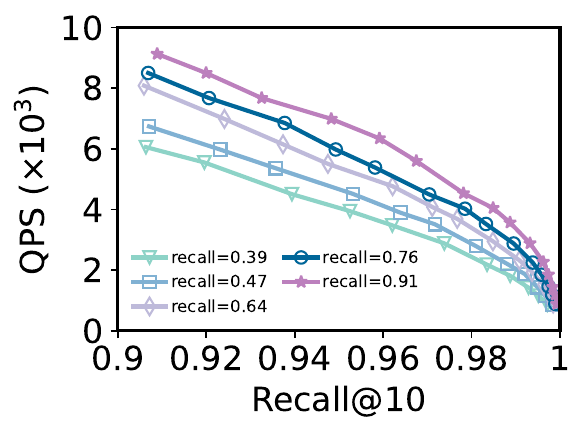}
        \label{fig:recall_sp_sift}
    }
    \subfigure[HNSW on \kw{Sift1M}]
    {
        \includegraphics[width=0.47\linewidth]{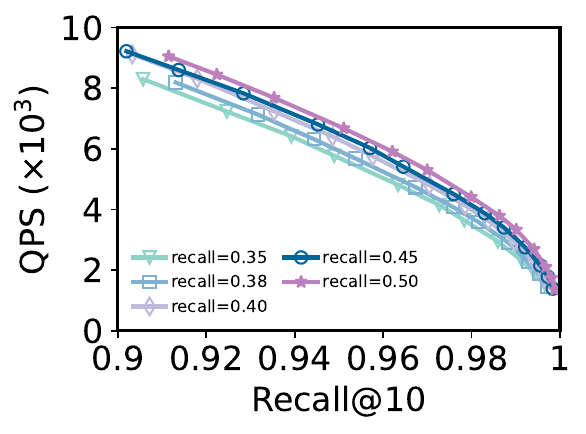}
        \label{fig:sift-qps-recall-hnsw}
    }
    \vspace{-3mm}
  \caption{The effects of the $k$-CNA quality (average recall of every node) on the search performance of the derived PG}
    \vspace{-3mm}
  \label{fig:recall_sp}
\end{figure}

\vspace{1mm}
\stitle{The Importance of $k$-CNA Quality:}
For the PG derived from $k$-CNA results by \kw{refinement} phase, there are two aspects of its search performance, i.e., efficiency by queries per second (QPS) and accuracy by $Recall@10$ of returned $k$-ANN. As depicted in Figure~\ref{fig:recall_sp}, each curve represents the search performance of a derived PG with a distinct recall of $k$-CNA results. 
The results clearly indicate that the $k$-CNA quality significantly impacts the search performance of the derived PG. Specifically, the recall of $k$-CNA results exhibits a positive effect on the search performance of PG. Such an effect could also observed on other PGs such as DPG and $\tau$-MNG. Notably, the $k$-CNA quality of HNSW is limited to 0.5 due to its incremental node-by-node insertion strategy of index construction. 

Considering the strong relationship between the quality of $k$-CNA and search performance, we identify two issues in the construction of RNG and NSWG respectively. 

\begin{figure}[t]
    \centering 
    \subfigure[Sift1M]
    {
        \includegraphics[width=0.42\linewidth]{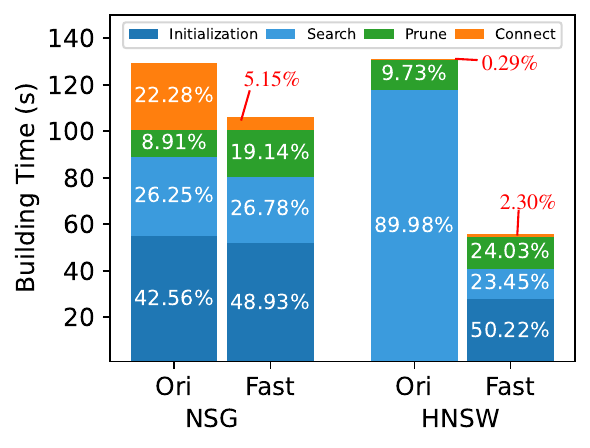}
        \label{fig:nsg_cost_decomp_nsg}
    }
    \subfigure[Gist1M]
    {
        \includegraphics[width=0.42\linewidth]{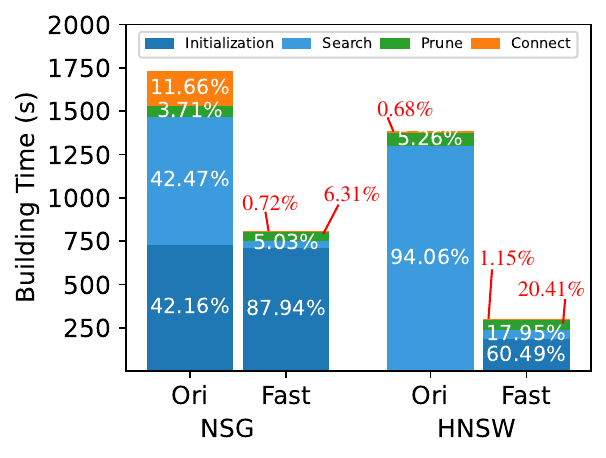}
        \label{fig:hnsw_cost_decomp_hnsw}
    }
    \vspace{-3mm}
    \caption{The cost decomposition of PG construction}\label{fig:cost_decomp_hnsw}
    \vspace{-3mm}
\end{figure}

\vspace{1mm}
\stitle{RNG Construction Issue:}
\emph{RNG methods suffer from inefficiency in obtaining $k$-CNA results}, caused by \kw{search} phase that takes the initial KNNG generated by \kw{initialization} phase as the graph index for accuracy improvement.
As demonstrated in previous experimental studies \cite{Dpg, survey2021}, with similar (e.g. tens of ) average out-degrees, KNNG is more prone to local optima than RNG and NSWG due to directional edges and weak connectivity. To address this, RNG equips the initial KNNG with a pretty large (e.g. hundreds of) out-degree, which enhances the $k$-CNA quality but leads to an inefficiency issue. As depicted in Figure~\ref{fig:cost_decomp_hnsw}, the search on KNNG to enhance the $k$-CNA quality constitutes a significant portion of the overall cost of building the representative RNG method (i.e., NSG), which even surpasses that taken for \kw{initialization} on the \kw{Gist1M} dataset. A similar phenomenon could be found in other RNG methods such as $\tau$-MNG. 


\vspace{1mm} 
\stitle{NSWG Construction Issue:}
\emph{NSWG methods suffer from a poor $k$-CNA quality in construction}, caused by its building strategy of incremental node-by-node insertions: utilizes the current graph index with only a part of nodes to conduct $k$-ANN search for $k$-CNA results. Hence, the expected value of $k$-CNA quality in NSWG is only $0.5$ even if all the $k$-ANN queries are answered correctly.
Since achieving exact correctness in $k$-ANN queries is not feasible, the average recall in practice is upper bounded by $0.5$, as demonstrated in Figure~\ref{fig:sift-qps-recall-hnsw}.



\section{Refinement Before Search}
\label{sec:new-strategy}

In this section, we focus on addressing the issue identified in the last section regarding the RNG construction. At a high level, we propose replacing the \emph{search-before-refinement} scheme (Figure~\ref{fig:build_nsg}) with a \emph{refinement-before-search} scheme (Figure \ref{fig:build_pg}) in RNG construction.
To enhance the efficiency of acquiring high-quality $k$-CNA results, we introduce a novel pruning strategy, $\alpha$-pruning for neighbor selection in \kw{refinement} (Section~\ref{sec:alpha}). Then, we theoretically analyze our proposed scheme to demonstrate its efficacy (Section~\ref{sec:exist-prune}).

\subsection{$\alpha$-Pruning Strategy}
\label{sec:alpha}

To enhance the $k$-CNA efficiency in RNG, we do not use the initial KNNG $G$ as the index used in \kw{search} phase due to its large out-degree. In our new scheme, we conduct the first \kw{refinement} on KNNG $G$ to obtain an RNG index denoted as $\hat{G}$ with a much smaller out-degree, then conduct the \kw{search} phase on $\hat{G}$ to obtain the $k$-CNA results, which are further used to produce the final RNG index via the second \kw{refinement}.

However, altering the order of \kw{refinement} and \kw{search} directly is unsuitable for index construction due to the following issue.

\vspace{1mm}
\stitle{RNG Pruning Issue:}
In Section~\ref{sec:rng}, we discuss the RNG pruning strategy used in the \kw{refinement} as shown in Figure~\ref{fig:example-orune}, the edge $(u,v)$ will be pruned if there exists edge $(u,w)$ such that $dist(u,w)<dist(u,v)$ and $dist(v,w)<dist(u,v)$.
However, as shown in the following example, it is primarily not designed for $k$-CNA.

\begin{example}
As shown in Figure~\ref{fig:example-orune}, $w$ and $v$ are out-neighbors of $u$ in KNNG, and the RNG pruning strategy applied leads to the pruning of edge $(u,v)$ by $w$. Consider a scenario where there exists a query point $q$ such that $dist(u,q) < dist(v,q) < dist(w,q)$.
During beam search on KNNG, $v$ is found when $u$ is included in $pool$ as defined in Algorithm~\ref{alg:knn_search}. However, after pruning, $v$ may no longer be found even if $u$ is included in $pool$. This is because $w$ might not be successfully inserted in $pool$ due to its longer distance. 
Thus, in such cases, the $k$-CNA quality on NSG is inferior to that on KNNG.
\label{rng-example}
\end{example}

Notably, the RNG pruning is designed to guarantee the finding of the $1$-NN through greedy routing when $q \in D$~\cite{Nsg}. Following this approach, several other pruning strategies have been introduced, e.g., \cite{taumg,Nssg,diskann}.
However, to the best of our knowledge, there is currently no existing study that specifically addresses the pruning strategy for $k$-CNA beyond the context of finding the $1$-NN.

Motivated by this, we introduce a novel pruning strategy named $\alpha$-pruning for the \kw{refinement}. Different to RNG pruning strategy, our strategy enables efficient retrieval of $k$-CNA results while allowing the control of the $k$-CNA quality by the parameter $\alpha$.

\vspace{1mm}
\stitle{$\alpha$-pruning:}
{An edge $(u,v)$ exists in the graph only if there is no edge $(u,w)$ in the graph where $dist(u,w) < dist(u,v)$, $dist(v,w) < dist(u,v)$, and $\angle uwv > \alpha$.}
\label{sec:alpha}

The practical effectiveness of our proposed $\alpha$-pruning is demonstrated in the following example, with theoretical analysis to follow in the subsequent subsection.

\begin{figure}
  \centering
    \subfigure[Recall v.s. Beam Width]
    {
        \includegraphics[width=0.47\linewidth]{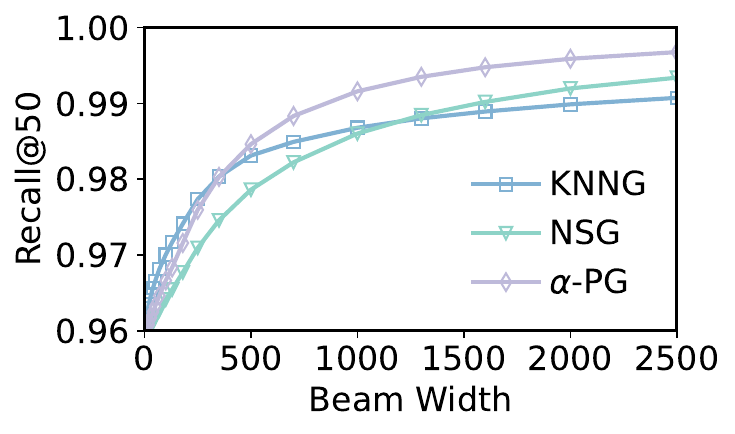}
        \label{fig:recall-width}
    }
    \subfigure[Recall v.s. Time]
    {
        \includegraphics[width=0.47\linewidth]{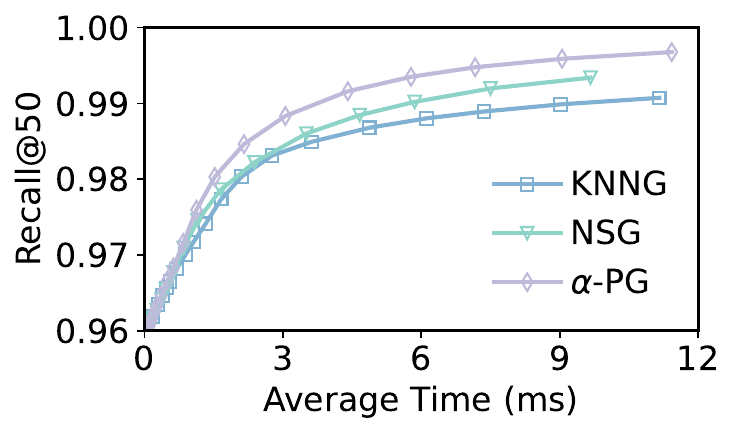}
        \label{fig:recall-time}
    }
    \vspace{-3mm}
  \caption{Comparing $k$-CNA results on Glove.}
  \vspace{-3mm}
  \label{fig:prune-loss}
\end{figure}

\begin{example}
To compare the $k$-CNA results, we randomly sample query points from the dataset as queries, and conduct $50$-ANN search with varying beam widths on \kw{Glove} dataset using three different graph indexes: KNNG, NSG, and $\alpha$-PG with $\alpha=66^\circ$. The accuracy of $k$-ANN search results is assessed from two angles: comparing results with the same beam width and comparing results within the same running time. 
In Figure~\ref{fig:prune-loss}, with the same beam width setting, it is evident that compared to KNNG, NSG compromises the quality of $k$-CNA results, whereas $\alpha$-PG maintains similar or superior quality, especially with increasing beam width. From a time-based perspective, the search on $\alpha$-PG emerges as the most effective choice.
\end{example}

\subsection{Analysis of Our Scheme}
\label{sec:exist-prune}


In this subsection, we evaluate the performance of the \emph{refinement-before-search} scheme by assessing the quality loss incurred by our pruning strategies in comparison to directly searching on KNNG.

Our discussion is based on comparing the search paths from the query to the ground-truth nodes on $G$ and $\hat{G}$. 

To be formal, we denote the KNNG as $G$ and the one obtained by applying $\alpha$-pruning on $G$ as $\hat{G}$.
For a $k$-CNA query $q$, we sort all nodes based on their distance to $q$, we assign them ranks denoted as $p_1, p_2, \ldots, p_n$, where the subscript represents the rank of each node, i.e., $p_i$ is the $i$-NN of $q$ and $rank(q, p_i) = i$.
For a path $P=[v_1,\cdots, v_m]$, we denote $\delta(P)=\max_{x\in \{v_1,\cdots, v_m\}}rank(q,x)$, that is, $\delta(P)$ denotes the maximum rank of nodes among path $P$.
For each node $u \in V(G)$, let $SP(q, u)$ be the set containing all paths from $q$ to $u$ in graph $G$, we use $\Delta(G,q,u)$ to denote the path in $SP(q,u)$ that minimizes the maximum rank among its nodes, i.e., $\Delta(G,q,u)=\argmin_{P\in SP(q,u)}\delta(P)$.

\begin{theorem}
\label{tm:rank}
Assume $S_1,S_2$ are the results of beam search with width $L$ for query $q$ on KNNG $G$ and pruned KNNG $\hat{G}$ respectively.
If $\delta(\Delta(G,q,p_k))=\delta(\Delta(\hat{G},q,p_k))$, we have $p_k\in S_1$ implies $p_k\in S_2$.
\end{theorem}

\proofsketch
We assume $\Delta({G},q,p_k)$ and $\Delta(\hat{G},q,p_k)$ are distinct; otherwise, the proof is trivial.
We prove it by contradiction.
Assume $p_k\in S_1,p_k\not\in S_2$ when $\delta(\Delta(G,q,p_k))=\delta(\Delta(\hat{G},q,p_k))$.
Then, there must exist a node $v\in \Delta(\hat{G},q,p_k)$ that has not been successfully inserted into the beam search queue, indicating that the path from $q$ to $p_k$ has not been discovered. Since $\Delta(\hat{G},q,p_k)$ exists in $G$ and does not disturb the search on path $\Delta({G},q,p_k)$, we have $rank(v,q)<\delta(\Delta(\hat{G},q,p_k))$.
However, for the same reason, such a node $v$ does not exist, because the node with $\delta(\Delta({G},q,p_k))$ has been successfully inserted, not to mention a node like $v$ with a lower rank. Hence, such $v$ does not exist, which leads to a contradiction. 
\eop

According to Theorem~\ref{tm:rank}, a node with a higher rank appearing in the search path on $\hat{G}$ indicates that we miss a closer neighbor in the $k$-CNA results, i.e., $k$-CNA quality loss compared with the search on $G$.
Such a loss is caused by the $\alpha$-pruning operations on $G$ for the sake of $k$-CNA efficiency. 
Being a widely-used pruning strategy, we have demonstrated in Example~\ref{rng-example} that RNG pruning results in higher ranks in the path, leading to quality loss. Next, our focus shifts to analyzing our newly proposed $\alpha$-pruning strategy.

\begin{figure}[t]
      \centering
      \begin{minipage}{0.49\linewidth}
    \includegraphics[width=0.93\columnwidth]{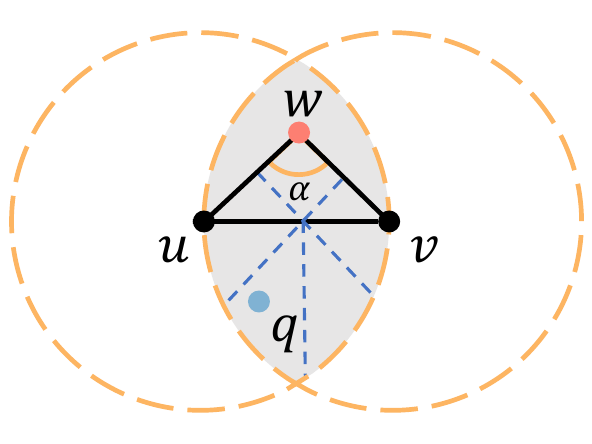}
    \caption{An example of $\alpha$-pruning strategy}
\label{fig:example-orune}
      \end{minipage}
      \hfill
      \begin{minipage}{0.48\linewidth}
        \includegraphics[width=0.9\columnwidth]{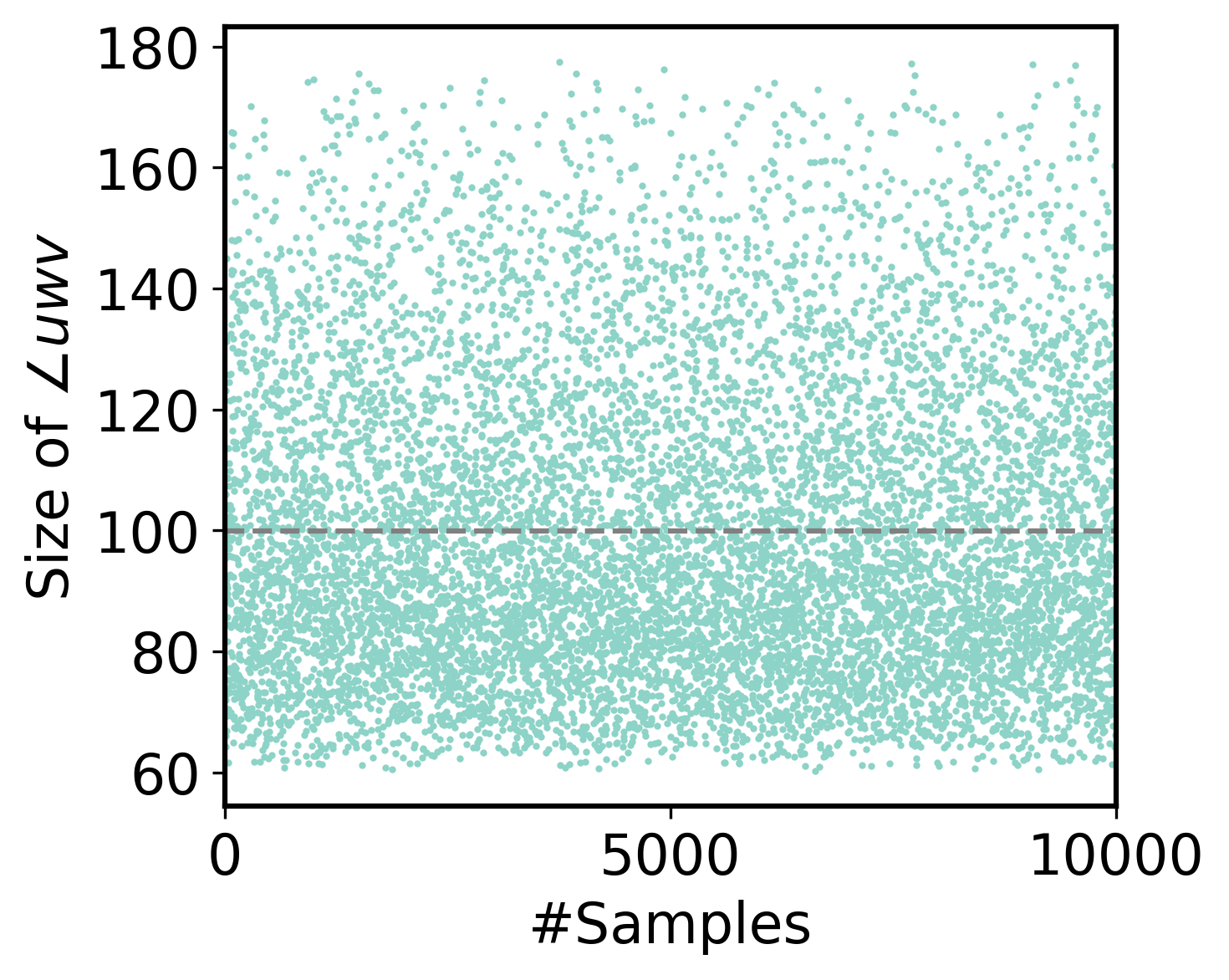}
        \caption{The size of $\angle uwv$ by Monte Carlo simulations}
\label{fig:angle}
      \end{minipage}
      \vspace{-3mm}
\end{figure}

We employ the following lemma to establish the relationship between $\alpha$ and the $k$-CNA quality loss caused by $\alpha$-pruning. 

\begin{lemma}
\label{lemma:prune-pro}
Assume the points are uniformly distributed in infinite space.
If the RNG pruning strategy prunes an edge $(u,v)$ caused by the edge $(u,w)$, and $\angle uwv=\alpha$, then the probability that $w$ has a higher rank than both $u$ and $v$ for any query point is $\frac{\pi-\alpha}{2\pi}$.
\end{lemma}

\proofsketch
Since the distance from $w$ to query $q$ is greater than the distance from $u$ and $v$, it implies that $q$ must be located in the region divided by the perpendicular hyperplane between $uw$ and $vw$ and farthest from $w$. The angle corresponding to that region is $\pi-\alpha$. Therefore, there is a probability of $\frac{\pi-\alpha}{2\pi}$ that $w$ has a higher rank than both $u$ and $v$ for query $q$.
\eop

According to Lemma~\ref{lemma:prune-pro}, as $\alpha$ increases, the probability that $\alpha$-pruning leads to $k$-CNA quality loss decreases. 
As shown in Figure~\ref{fig:angle}, through Monte Carlo simulations, the expected value of the angle $\angle uwv$ is approximately $100^\circ$.
This indicates that one successful $\alpha$-pruning leads to a higher maximum rank along the path (i.e., $k$-CNA quality loss) with about $0.2$ $(\frac{180-100}{360})$ probability. 
Hence, we can control such a probability via $\alpha$. 
Lemma~\ref{lemma:prune-pro} describes the relationship between $\alpha$ and $k$-CNA quality loss in the perspective of the neighborhood of a single node. In the following, we further present such a relationship in the view of the search path on $\hat{G}$.



\begin{theorem}
Assume the points are uniformly distributed in infinite space.
For any path $P=[q,\cdots,p_k]$, at least successfully utilizing $\alpha$-pruning strategy $\frac{2\pi}{\pi-\alpha}$ times in expectation leads to a larger value $\delta(P)$ of the path.
\label{theo:expected}
\end{theorem}

\proofsketch
From lemma~\ref{lemma:prune-pro}, we know the successful $\alpha$-pruning has at most $Pr=\frac{\pi-\alpha}{2\pi}$ probability leads to a higher rank. Since each $\alpha$-pruning is independent, the expected number of $\alpha$-pruning leads to a higher rank is $1Pr+2Pr(1-Pr)+3Pr(1-Pr)^2+\cdots=1/Pr.$
Hence, at least $\frac{2\pi}{\pi-\alpha}$ $\alpha$-pruning in expectation leads to a higher rank in the path.
\eop


According to Theorem~\ref{theo:expected}, when $\alpha$ is small, a short search path from $q$ to $p_k$ will lead to $k$-CNA quality loss in expectation. For example, when $\alpha = 60^\circ$, $k$-CNA quality loss happens once the path length reaches 3 in expectation. However, as we are aware, the path length from $u$ to $p_k$ is approximately $6$~\cite{travers1977experimental,Dpg}.
Hence, this could result in quality loss when applying the RNG pruning strategy directly. Fortunately, through our proposed $\alpha$-pruning strategy, we can mitigate the risk of quality loss in $k$-CNA by carefully setting the value of $\alpha$.
Note that when $\alpha = 60^\circ$, $\alpha$-pruning equals the RNG pruning. Hence, by setting $\alpha > 60^\circ$, $\alpha$-pruning is able to reduce the $k$-CNA quality loss caused by our \emph{refinement-before-search} scheme compared with the RNG pruning. We will discuss the selection of $\alpha$ in the exp.1-b in Section~\ref{sec:exp}.

\section{A New PG Construction Framework}
\label{sec:new-framework}

In this section, we provide comprehensive details of our new construction framework for RNG and NSWG. 
Combining the approaches outlined in the previous section, we first present an optimized $k$-CNA approach in Section~\ref{sec:search-opt}, and then we present our new construction methods for RNG on top of our optimized $k$-CNA approach in Section~\ref{sec:iteration}. 
Further, we enhance the NSWG construction by combining a layer-by-layer insertion strategy with the RNG construction framework in Section~\ref{sec:alg-hii}.
In Section~\ref{sec:impl}, optimization techniques are introduced to enhance the efficiency of RNG construction framework. 
Lastly, we consolidate all the methods discussed and present a streamlined and effective framework that can be applied to the construction of other PG methods, as in Section~\ref{sec:summary}.

\subsection{Optimized $k$-CNA Approach}
\label{sec:search-opt}

In this part, we present the details of our $k$-CNA method following the \emph{refinement-before-search} scheme proposed in the last section. 
We present OptKCNA in Algorithm~\ref{alg:opt_build_knng}. 
It takes the current $k$-CNA results $\{ C(u) | u \in D \}$ as input, where $C(u)$ could be the $k$ out-neighbors from the initial KNNG $G_{k_0}$, and outputs the refined $k$-CNA results by two steps, i.e. \kw{refinement} and \kw{search}. 
In \kw{refinement}, a PG index $\hat{G}$ is directly derived by first applying $\alpha$-pruning on $G_{k_0}$ (Lines 1-4), where $C(u) = N_{G_{k_0}}(u)$, and then enhancing the graph connectivity (Lines 5-6). 
In \kw{search}, the $k$-CNA results of each node $u$ is enhanced via a $k$-ANN search on $\hat{G}$ (Lines 7-8).

Compared with the RNG construction, our method can be seen as a reversal of \kw{search} and \kw{refinement}.
However, OptKCNA is more efficient due to the smaller node out-degrees and keeps the $k$-CNA quality with a high probability as our analysis in Section~\ref{sec:exist-prune}.


\begin{algorithm}[t]
\small  
    \SetVline 
    \SetFuncSty{textsf}
    \SetArgSty{textsf}
\caption{\texttt{OptKCNA}($\{C(u) | u \in D\}, k,L,M,\alpha$)}
\label{alg:opt_build_knng}
\Input{$\{C(u) | u \in D\}$ and four parameters $k$, $L$, $m$ and $\alpha$}
\Output{refined $k$-CNA results $\{C(u) | u \in D\}$}
\tcc{\textbf{Phase 2:} \texttt{Refinement} ({ $\hat{G} =$\kw{Refine}$(\{C(u) | u \in D\}, M, \alpha)$})}
\For{each $u \in D$ \underline{in parallel}}
{
\State{$N_{\hat{G}}(u)\leftarrow$ \kw{Prune}($u, C(u)$, M, $\alpha$)}
}
\For{each $u \in D$ \underline{in parallel}}
{
\State{$N_{\hat{G}}(u)\leftarrow$ \kw{Prune}($u, N_{\hat{G}}(u) \cup \{v|u\in N_{\hat{G}}(v)\}, M, \alpha$)}
}
\State{find connected components via DFS}
\State{add extra edges into $E(\hat{G})$ between connected components}
\tcc{\textbf{Phase 3:} \texttt{Search}}
\For{each $u \in D$ \underline{in parallel}}
{
    \State{$C(u)\leftarrow$ \texttt{KANNSearch} ($\hat{G}, u, k, L, u$)}
}
\Return{$\{C(u) | u \in D\}$}
\end{algorithm}

\subsection{Self-Iterative Construction of RNG}
\label{sec:iteration}

According to our \emph{refinement-before-search} scheme, we could generate the final RNG index by applying another \kw{refinement} on the results of OptKCNA. 
However, such a simple construction strategy still faces the following challenges, in order to balance the $k$-CNA efficiency and quality.

\vspace{1mm}
\stitle{Difficulty of Tuning Parameters in \kw{Search} Phase:}
In the \kw{search} phase, $L$ is the key parameter to the efficiency-quality balance and thus should be carefully selected. However, the best choice of $L$ varies from datasets. Grid search is commonly employed for finding the best $L$~\cite{survey2021,Dpg}, which builds a corresponding graph index for each potential $L$ value with a huge cost. Hence, this one-shot strategy of setting $L$ is considerably time-consuming. 

To address this issue, we take the progressive strategy of tuning $L$ and propose a self-iterative framework with two key components, i.e., \textbf{quality examination} and \textbf{iterative refinement}.
The former determines the termination condition of our framework, while the latter defines the behaviour in each iteration. Again, we use the NSG as the representative RNG to illustrate the details.


\vspace{1mm}
\stitle{Quality Examination:}
As determining the quality of NSG itself is challenging, we utilize the $k$-CNA quality to assess the quality of the derived NSG. However, computing the $k$-CNA quality can be time-consuming, as it requires brute-force computation of the ground truth (exact $k$-nearest neighbors of each node). To address this issue, we estimate the $k$-CNA quality via sampling. Specifically, we employ a random selection process to choose a specific number $n_s$ of nodes (we will discuss later) and then compute the average recall over the $k$-CNA results of those sampled nodes as an estimator for the average recall over all nodes, i.e., the $k$-CNA quality.

We then delve into determining the value of $n_s$ to ensure that the estimation obtained through random sampling possesses a theoretical guarantee.
Let $r(u)$ denote the quality of current $k$-CNA results of $u$, and $r(G) = \sum_{u \in G} r(u)$ the precise sum of recall of each node in the graph $G$.
Suppose we randomly select $n_s$ nodes $S = \{u_1, \cdots, u_{n_s}\}$. We can compute their sum of recall, denoted as $\hat{r}(S) = \sum_{u \in S} r(u)$.
By the Chernoff bounds~\cite{MotwaniR95}, the following theorem proves $\frac{\hat{r}(S)}{n_s}$ is an accurate estimator of $\frac{r(G)}{n}$ when the number of samples $n_s$ is sufficiently large.

\begin{algorithm}[t]
\small  
    \SetVline 
    \SetFuncSty{textsf}
    \SetArgSty{textsf}
\caption{\texttt{IterNSG}($D, k_0, k, L, M, \alpha$)}
\label{alg:iter_nsg}
\Input{dataset $D$ and five parameters $k_0$, $k$, $L$, $M$ and $\alpha$}
\Output{an NSG $G$}
\tcc{\textbf{Phase 1:} \texttt{Initialization}}
\State{build the initial KNNG $G_{k_0}$ via KGraph \cite{KGraph}}
\For{each $u \in D$  \underline{in parallel}}{
    \State{$C(u) = N_{G_{k_0}}(u)$}
}
\tcc{\textbf{Phase 4:} \texttt{Quality Examination}}
\While{the estimator $\hat{r}$ does not achieve the requirement}
{
    \State{$\{C(u) | u \in D\} \leftarrow$ \kw{OptKCNA}($\{C(u) | u \in D\},k,L,M,\alpha$)}
    \State{estimate the quality of $\{C(u) | u \in D\}$ as $\hat{r}$}
}
\tcc{\textbf{Phase 2:} \texttt{Refinement}}
\State{$G =$\kw{Refine}($\{C(u) | u \in D\}, M$) as Lines 5-10 in Algorithm~\ref{alg:build_nsg}}
\Return{$G$}
\end{algorithm}

\begin{theorem}
Assume that $n_s$ satisfies $n_s\ge {(8+2\varepsilon)l\log n}/{\varepsilon^2}$. Then the inequality $|\frac{\hat{r}(S)}{n_s}-\frac{r(G)}{n}|< \frac{\varepsilon}{2}$ holds with at least $1-n^{-l}$ probability.
\label{theo:eps}
\end{theorem}

\proofsketch
We regard $\hat{r}(S)$ as the sum of $n_s$ i.i.d. Bernoulli variables with a mean $\mu=r(G)/n$. Then we have $Pr[|\frac{\hat{r}(S)}{n_s}-\frac{r(G)}{n}|\ge \frac{\varepsilon}{2}]=Pr[|\hat{r}(S)-n_s\mu|\ge \frac{\varepsilon n_s}{2}]=Pr[|\hat{r}(S)-n_s\mu|\ge \frac{\varepsilon}{2\mu}\cdot n_s\mu]$. Let $\delta=\frac{\varepsilon}{2\mu}$, by the Chernoff bounds, $\mu\le 1$ and $n_s\ge \frac{(8+2\varepsilon)l\log n}{\varepsilon^2}$, we have $Pr[|\frac{\hat{r}(S)}{n_s}-\frac{r(G)}{n}|\ge \frac{\varepsilon}{2}]\le exp(-\frac{\delta}{2+\delta}\cdot n_s\mu)=exp(-\frac{\varepsilon^2n_s}{8\mu+2\varepsilon})\le \frac{1}{n^l}$.
\eop

It is worth noting that such estimation can be executed asynchronously with index construction, allowing for the evaluation of the index after each iteration while progressively building the index. Simultaneously, we can estimate $k$-CNA quality and terminate the process once deemed satisfactory.

\vspace{1mm}
\stitle{Iterative Refinement:}
Once we have the quality examination, we can design an iterative refinement approach for NSG construction. This involves conducting further searches on $\alpha$-PG (i.e., the graph index derived by $\alpha$-pruning on current $k$-CNA results with connectivity enhancement) when the quality is deemed insufficient. The details are presented in Algorithm~\ref{alg:iter_nsg}.
The \kw{initialization} phase remains the same as the original NSG, where $k$-CNA results are obtained from a KNNG (Line 1-3). 
The iterative process continues until the quality requirement is met (Line 4). Alternatively, the termination condition can be set based on the number of iterations.
Within each iteration, $k$-CNA results are refined using Algorithm~\ref{alg:opt_build_knng} (Line 5). Once the iterative process terminates, the $k$-CNA results are further pruned via RNG pruning and connectivity enhancement to obtain the NSG (Line 7).

\subsection{Global Construction of NSWG}
\label{sec:alg-hii}

\begin{algorithm}[t]
\small  
    \SetVline 
    \SetFuncSty{textsf}
    \SetArgSty{textsf}
\caption{\texttt{OptHNSW}($D,k_0,ef,M,\alpha$)}
\label{alg:opt_hnsw}
\Input{dataset $D$ and four parameters $k_0$, $ef$, $M$ and $\alpha$}
\Output{an HNSW $G$}
\For{each $u\in D$ \underline{in parallel}}
{
    \State{randomly determine the highest layer of $u$ as $l(u)$}
}
\State{$m_L \leftarrow \max_{u\in D}l(u)$}
\State{randomly select $ep$ from the points in layer $m_L$}
\For{$i\leftarrow m_L$ downto $0$ \underline{in parallel}}
{
    \State{$D_i\leftarrow \{u|l(u)\ge i\}$}
    \If{$|D_i|\le M$}
    {
        \State{$N_{G_i}(u)\leftarrow D_i$ \textbf{for} each $u\in D_i$}
    }
    \Else{
        \State{$G_i\leftarrow$\kw{IterNSG}($D_i,k_0,ef,ef,M, \alpha$)}
    }
}
\Return{$G = \{G_0, G_1, \ldots, G_{m_L}\}$}
\end{algorithm}

As mentioned in the previous section regarding the NSWG construction issue, the poor $k$-CNA quality primarily arises due to the incremental node-by-node insertion strategy and the \kw{search} conducted on the current incomplete graph index that includes only a portion of nodes. In this part, we still use HNSW as the representative NSWG index, and propose a global construction of HNSW, which achieves $k$-CNA results on whole data points in each layer instead of only a part.



Our global construction of HNSW is built on top of a layer-by-layer insertion strategy in a top-down manner.
To be specific, we first determine the layers for each node before the actual layer insertion and thus we obtain the whole set of nodes in each layer. Afterward, we employ our RNG construction framework in Section~\ref{sec:iteration} to build a graph index for each layer. In this way, each layer of HNSW is built on top of $k$-CNA results w.r.t the whole set of nodes in each layer instead of only a subset. Hence, the $k$-CNA quality of each layer will significantly exceed 0.5 (i.e., the upper bound in the original HNSW). Note that each layer of HNSW is actually an RNG index, since it takes the RNG pruning strategy to prune the $k$-CNA results only without the connectivity enhancement via DFS. Subsequently, we insert the RNG index of the layer into HNSW according to the layer-by-layer insertion strategy.

The details of our HNSW construction are presented in Algorithm~\ref{alg:opt_hnsw}. Initially, we randomly determine the layer of each node, following the same approach as in the original HNSW (Lines 1-3). Then, we select one node at the top layer as the entry node $ep$ (Line 4). The index construction proceeds from the top layer to the bottom layer (Line 5). For nodes in each layer (Line 6), we directly connect them if the number of nodes does not exceed the out-degree limit $M$ (Lines 7-8). Alternatively, we apply our optimized NSG construction to obtain the index (Lines 9-10).


\subsection{Implementation Details}
\label{sec:impl}

In this part, we discuss two crucial optimization techniques aimed at improving the construction efficiency of our RNG framework as well as NSWG framework.
To discuss the details briefly later, in the $i$-th iteration of Algorithm~\ref{alg:iter_nsg},  $k$-CNA results are formed as a KNNG (of the dataset) denoted as $G^K_i$, and the $\alpha$-PG is denoted as $G^N_i$.

\begin{figure}
  \centering
  \subfigure[P2K]
    {
    \includegraphics[width=0.47\linewidth]{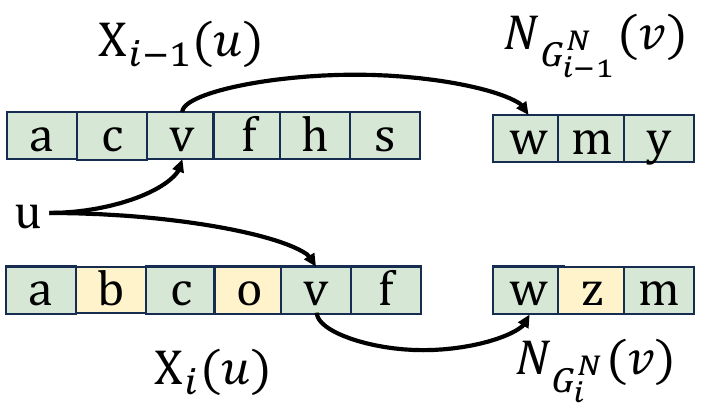}
    \label{fig:repetition_search}
    }
  \subfigure[K2P]
    {
     \includegraphics[width=0.47  \linewidth]{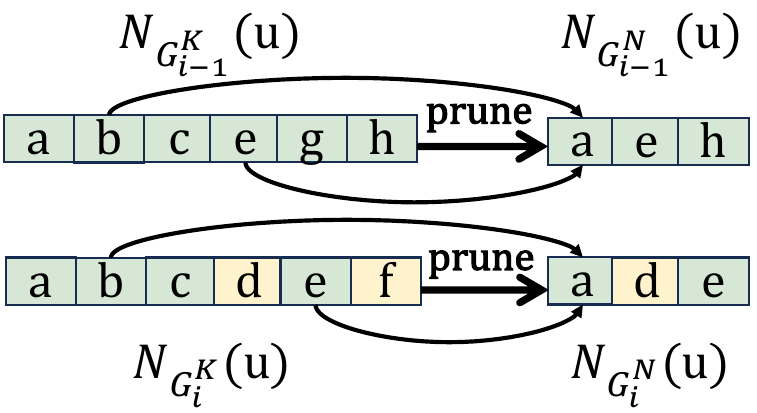}
    \label{fig:repetition_prune}
    }
    \vspace{-3mm}
  \caption{Examples of repeated distance computations}
  \vspace{-3mm}
  \label{fig:repetition}
\end{figure}

\vspace{1mm}
\stitle{\texttt{P2K}: from $\alpha$-PG to KNNG.}
This process entails answering a $k$-ANN query on the current $\alpha$-PG for each $u \in D$. The optimization efforts primarily avoid repeated distance computations. It is evident that multiple iterations of \texttt{P2K} involve repeated distance computations, as each point $u$ is inquired multiple times, leading to redundant verification of its similar points. This happens during node expansions, as depicted in Lines 5-6 of Algorithm \ref{alg:knn_search}, when conducting $k$-ANN search for node $u$. 

\begin{example}
Figure \ref{fig:repetition_search} shows an example, where the nodes colored yellow indicate new members in $i$-th iteration and $X_i(u)$ denotes the set of expanded nodes in the $i$-th iteration. In the example, the search path $u\rightarrow v\rightarrow w$ is found in two consecutive iterations and thus $dist(u,w)$ are computed twice.
\end{example}

In \texttt{P2K}, the computation of $dist(u, w)$ occurs in both the $(i-1)$-th and $i$-th iterations, when $v \in X_{i-1}(u) \cap X_{i}(u)$ and $w \in N_{G^N_{i-1}}(v) \cap N_{G^N_{i}}(v)$. In this case, $v$ is expanded and $w$ is a neighbor of $v$ in both iterations. Furthermore, it is worth noting that for each node $v \in D$, $N_{G^N_i}(v)$ remains similar (with only a small portion being changed) when $N_{G^K_{i-1}}(v)$ is sufficiently accurate. As a result, the size of $N_{G^N_{i-1}}(v) \cap N_{G^N_{i}}(v)$ increases resulting in more repeated distance computations.

To address this issue, we propose adding additional information to the neighbor list of each node. Firstly, we compare $X_{i}(u)$ and $X_{i-1}(u)$ to determine whether each $v \in X_{i}(u)$ has already been in $X_{i-1}(u)$. We accomplish this by adding a boolean value to each member in $X_{i}(u)$. Secondly, we compare $N_{G^N_{i-1}}(v)$ and $N_{G^N_{i}}(v)$ to record whether each $w \in N_{G^N_i}(v)$ has appeared in $N_{G^N_{i-1}}(v)$. Similarly, we assign a boolean value to each member in $N_{G^N_{i}}(v)$. If a vertex $v \in X_{i}(u) \setminus X_{i-1}(u)$, we consider $N_{G^N_{i}}(v)$ as candidates of $u$. Otherwise, we only consider $N_{G^N_{i}}(v) \setminus N_{G^N_{i-1}}(v)$ as candidates. By implementing this approach, we can effectively reduce the number of repeated distance computations in consecutive iterations of \texttt{P2K}.

\vspace{1mm}
\stitle{\texttt{K2P}: from KNNG to $\alpha$-PG.}
We discuss the optimization of \texttt{K2P}.
Like \texttt{P2K}, pruning operations for the same node are done in multiple iterations. Thus, there exist repeated distance computations and angle computations in consecutive iterations.

\begin{example}
As shown in Figure \ref{fig:repetition_prune}, let us consider $a \in N_{G^N_{i-1}}(u) \cap N_{G^N_i}(u)$ and $b \in N_{G^K_{i-1}}(u) \cap N_{G^K_i}(u)$. 
Notably, both $N_{G^N_i}(u)$ and $N_{G^K_i}(u)$ are sorted in ascending order of the distance from $u$. Hence, $dist (u, a) < dist (u, b)$ holds. 
In such case, $dist(a, b)$ and $\angle uab$ will be computed twice when pruning the edge set $\{(u,v)|v\in N_{G^K_{i-1}}(u))\}$ and $\{(u,v)|v\in N_{G^K_{i}}(u))\}$ respectively. 
That is because, when we check $b$ in both iterations, $a$ has been in $N_{G^N_i}(u)$ and we have to decide whether or not the edge $(u,a)$ dominates $(u,b)$.
\end{example}

Besides, as $i$ increases, the quality of $G^K_i$ improves, which results in an increase in unnecessary computations.
Therefore, to reduce such repetition, we take a strategy like that in \kw{P2K}. To be specific, we add extra information to distinguish whether or not each $b \in N_{G^K_i}(u)$ is also a member of $N_{G^K_{i-1}}(u) $ and each $a \in N_{G^N_i}(u)$ a member of $N_{G^N_{i-1}}(u)$. 
This could be easily implemented by adding an extra boolean value for each member in $N_{G^K_i}(u)$ and $N_{G^N_i}(u)$. 

With such information, let us consider whether $b \in N_{G^K_i}(u)$ will join $N_{G^N_i}(u)$. If $b \notin N_{G^K_{i-1}}(u)$, we deal with it as the normal pruning, since there exist no distance computations between $b$ and members in $N_{G^N_{i-1}}(u)$. Otherwise, we have $b \in N_{G^K_i}(u) \cap N_{G^K_{i-1}}(u)$. If $b \in N_{G^N_{i-1}}(u)$, we only check whether members in $N_{G^N_i}(u) \setminus N_{G^N_{i-1}}(u)$ but omit others in $N_{G^N_i}(u)$. Otherwise, we process it with a normal pruning. Hence, we can reduce those unnecessary computations without changing the final pruning results.

\subsection{Summary}
\label{sec:summary}

The pipeline of our framework is summarized in Figure~\ref{fig:build_pg}. In the beginning, we obtain a KNNG through the \kw{initialization} phase and then enter an iterative loop to continuously enhance the $k$-CNA quality until it satisfies the quality examination or achieves the required number of iterations. During each iteration, we first obtain an intermediate graph index from the current $k$-CNA results via a new pruning strategy for neighbor selection in the \kw{refinement} phase. Subsequently, we perform a beam search for each node in the \kw{search} phase to enhance the $k$-CNA quality. The optimization techniques, i.e., \kw{K2P} and \kw{P2K}, further accelerate the \kw{refinement} and \kw{search} phases respectively.

It is important to note that our framework is not only well-suited for NSG and HNSW, as detailed in this section, but also could be extended to the construction of other SOTA PGs, such as $\tau$-MNG~\cite{taumg} and NSW~\cite{Sw} (as demonstrated in Exp.5 in Section~\ref{sec:exp}). 
This is due to the fact that the construction pipelines of all SOTA PGs fall into the two categories we have discussed.

{

\noindent \textbf{GPU Implementation.}
Our framework can be implemented on GPU.
Firstly, in the initialization phase, we can utilize GNND~\cite{GNND}, a GPU-based state-of-the-art KNNG construction method, to replace KGraph, as both GNND and KGraph are followed by the idea of NN-Descent~\cite{KGraph}. 
Secondly, for the search phase, we can parallelize each search operation by treating each query independently. 
Thirdly, in the refinement phase,
(i) since pruning on each vertex is independent, it can be efficiently parallelized; 
(ii) for connecting, its key operation is to detect connected components in the graph, which can be solved using existing GPU-based solutions, e.g., \cite{AlabandiSBB23}.
}

\section{Experiments}
\label{sec:exp}

{In this section, we present the results of our experimental study. We begin by introducing the experimental settings, followed by showcasing the building cost and search performance results when applying our framework to NSG, HNSW, $\tau$-MNG and NSW. Furthermore, we compare our framework with four up-to-date methods to highlight the enhanced efficiency of our index construction without compromising search performance.
Lastly, we analyze the effects of the optimization techniques employed and demonstrate the scalability of our approach on a large dataset.

\stitle{Datasets:}
We use 6 public datasets with diverse sizes and dimensions. These datasets encompass a wide range of applications, including image (\texttt{Sift1M}~\cite{sift}}, \kw{Deep1M}~\cite{survey2021} and \texttt{Gist1M}~\cite{sift}), audio (\texttt{Msong}~\cite{msong}) and text(\texttt{Crawl}~\cite{crawl} and \kw{Glove}~\cite{glove}).
The statistics of those data sets are summarised in Table \ref{tb:data}, where \emph{\#queries} denotes the number of queries and \emph{dim.} denotes the dimensions of datasets.
The query workloads of the datasets are given in the datasets.
{Besides, we take several random samples of distinct sizes from Sift50M~\cite{sift} dataset to test the scalability of our methods.}

\begin{table}
\centering
\caption{Data statistics}
\vspace{-3mm}
\label{tb:data}
\begin{tabular}{|c|r|r|r|c|c|}
\hline
\textbf{Dataset} &  \textbf{size} &  \textbf{\#queries}&  \textbf{dim.}& \textbf{type} \\
\hline \hline
Sift1M & 1,000,000 &10,000& 128& Image\\
\hline
Gist1M & 1,000,000 &1,000& 960& Image \\
\hline
Msong & 992,272 &200& 420& Audio\\
\hline
Crawl & 1,989,995 &10,000& 300& Text \\
\hline
Glove & 1,183,514  & 10,000&100& Text \\
\hline
Deep1M & 1,000,000 &10,000& 96& Image \\
\hline
\end{tabular}
\vspace{-3mm}
\end{table}

\stitle{Performance Indicators:}
Given a PG construction method, we care about two aspects of performance, i.e., construction cost and search performance. We use the execution time to evaluate the construction cost, denoted as ``Building Time''. For search performance, we care about efficiency measured by queries per second (QPS) and accuracy evaluated by recall. Given a query $q$, let $k$ denote the number of returned neighbors, $N^*(q)$ be the exact $k$-nearest neighbors of $q$, while $N(q)$ be the set of $k$ returned neighbors from different algorithms. The recall of the returned result is defined as $Recall@k = |N^* (q) \cap N(q)| / k$. We use the average recall over the query set to estimate the accuracy. We set the number $k$ as $10$ by default unless specified. All results are averaged over 5 runs. 

\stitle{Computing Environment:}
{All experiments are conducted on a server equipped with 2 Intel(R) Xeon(R) Silver 4210R CPUs, each of which has 10 cores (each supporting 2 hyper-threads), and {256 GB DRAM} as the main memory. The OS version is CentOS 7.9.2009. All codes were written by {C++} and compiled by {g++ 11.3}. The SIMD instructions are enabled to accelerate the distance computations.}


\stitle{Compared Algorithms:}
We mainly consider comparing the construction processes of two representative PG-based methods, i.e., NSG and HNSW, with our newly proposed framework. We refer to the original construction methods of NSG and HNSW as \kw{OriNSG} and \kw{OriHNSW} respectively. Additionally, we introduce our framework to enhance the construction of NSG and HNSW, denoted as \kw{FastNSG} (Algorithm~\ref{alg:iter_nsg}) and \kw{FastHNSW} (Algorithm~\ref{alg:opt_hnsw}) respectively.
We further compare our construction framework with three recently developed PG construction algorithms: \kw{DiskANN}~\cite{diskann}, \kw{LSH}-\kw{APG}~\cite{Lsh-apg}, \kw{RNN}-\kw{Descent}~\cite{OnoM23} and ParlayANN \cite{ManoharSBD0S024}.
In order to assess the generality of our framework, we conduct performance comparisons on  {$\tau$-MNG}~\cite{taumg} and NSW~\cite{Sw}, both with (denoted as \kw{Fast}$\tau$-\kw{MNG} and \kw{FastNSW}) and without (denoted as \kw{Ori}$\tau$-\kw{MNG} and \kw{OriNSW}) the utilization of our proposed framework.

{The index construction comparisons use all threads on the server, while the searches are compared using one single thread.}

\begin{figure}[t]

  \centering
  \subfigure[QPS v.s. Recall]{
    \includegraphics[width=0.47\linewidth]{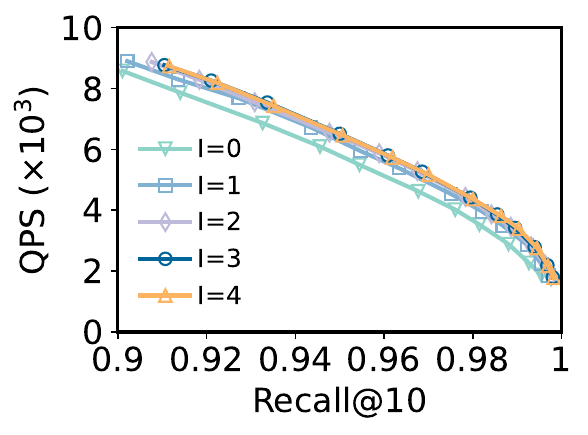}
    \label{fig:iter_search}
  }
  \hfill
  \subfigure[Building time]{
     \includegraphics[width=0.47\linewidth]{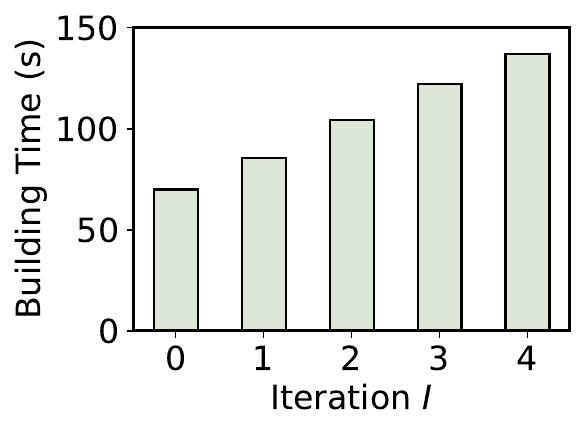}
    \label{fig:iter_build}
  }
  \vspace{-3mm}
  \caption{Effects of number of iterations on SIFT1M (Exp.1-a)}
  \vspace{-3mm}
  \label{fig:iter}
\end{figure}

\begin{figure}[t]

  \centering
  \subfigure[Glove]{
    \includegraphics[width=0.47\linewidth]{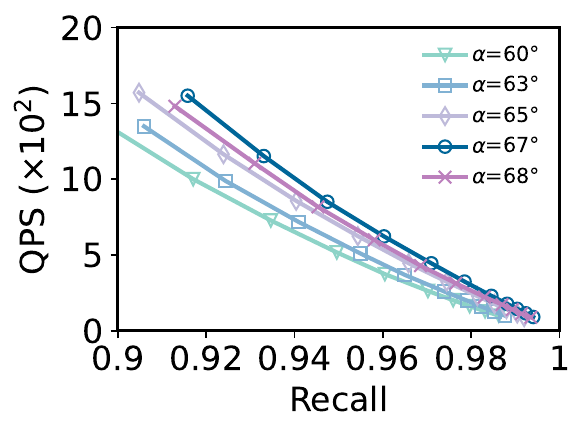}
    \label{fig:msong-alpha}
  }
  \hfill
  \subfigure[Msong]{
     \includegraphics[width=0.47\linewidth]{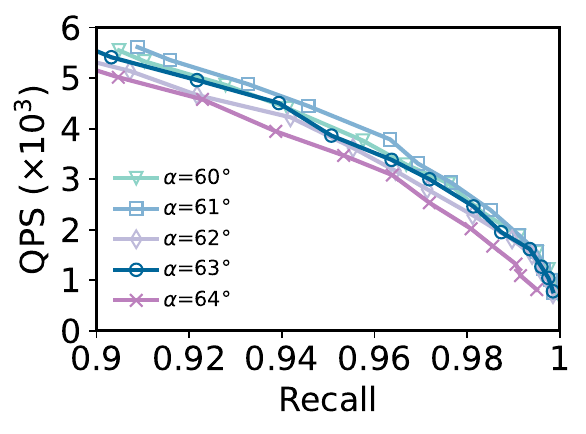}
    \label{fig:glove-alpha}
  }
  \vspace{-3mm}
  \caption{Effects of value of $\alpha$ (Exp.1-b)}
  \vspace{-3mm}
  \label{fig:alpha-exp}
\end{figure}


\begin{figure*}[t]
\includegraphics[width=0.35\linewidth]{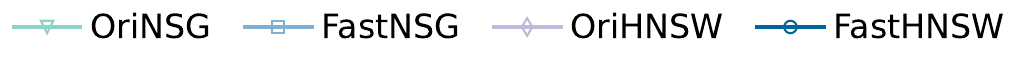}

  \centering
  \subfigure[Sift1M ($k=10$)]{
    \includegraphics[width=0.154\linewidth]{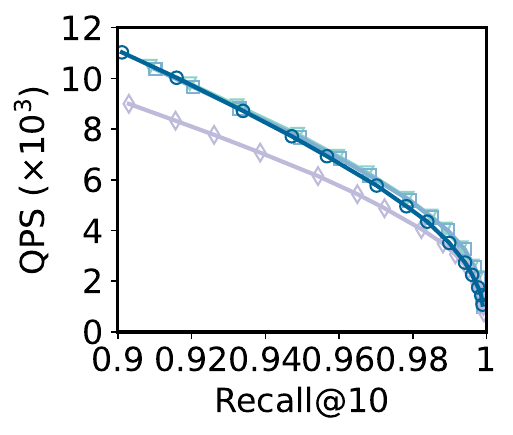}
    \label{fig:main_sift}
  }
  \hfill
  \subfigure[Gist1M ($k=10$)]{
     \includegraphics[width=0.154\linewidth]{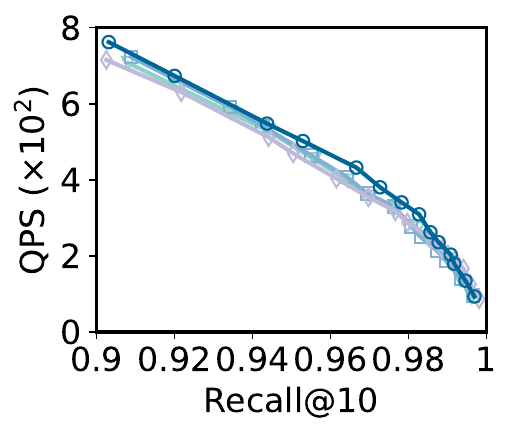}
    \label{fig:rmain_gist}
  }
  \hfill
  \subfigure[Msong ($k=10$)]{
    \includegraphics[width=0.154\linewidth]{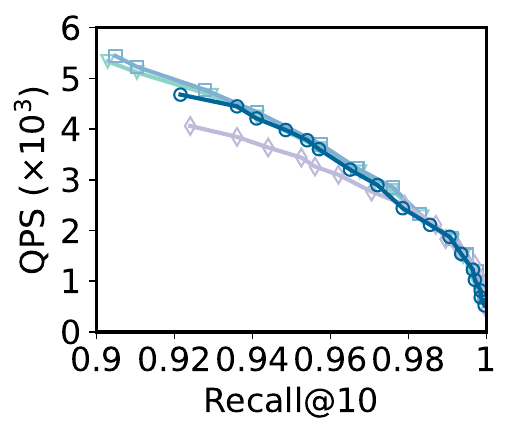}
    \label{fig:main_msong}
  }
  \hfill
  \subfigure[Crawl ($k=10$)]{
     \includegraphics[width=0.154\linewidth]{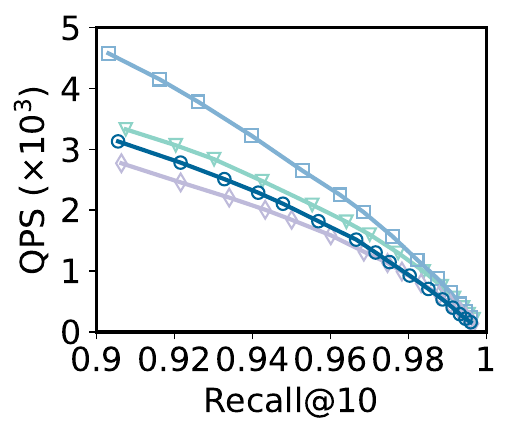}
    \label{fig:main_crawl}
  }
  \hfill
  \subfigure[Glove ($k=10$)]{
     \includegraphics[width=0.154\linewidth]{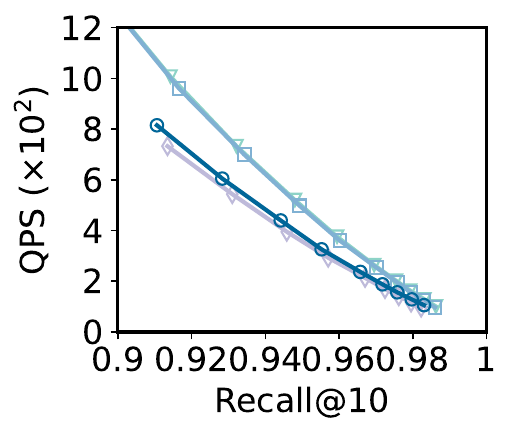}
    \label{fig:main_glove}
  }
  \hfill
  \subfigure[Deep1M ($k=10$)]{
     \includegraphics[width=0.154\linewidth]{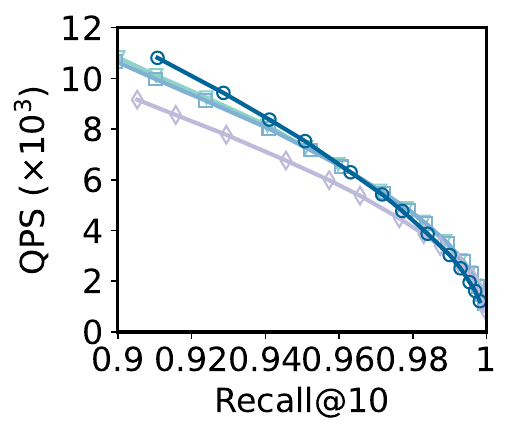}
    \label{fig:main_deep}
  }
   \hfill
   \subfigure[Sift1M ($k=15$)]{
      \includegraphics[width=0.154\linewidth]{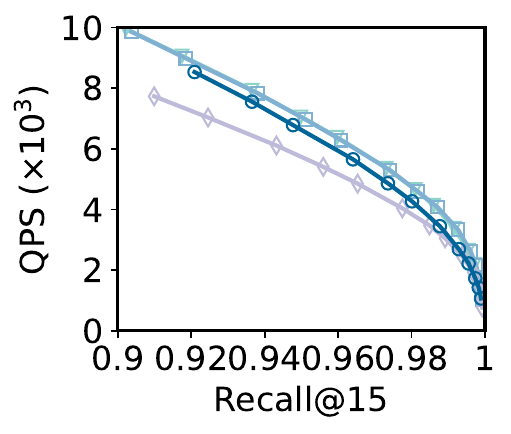}
     \label{fig:main_deep}
   }
   \hfill
   \subfigure[Gist1M ($k=15$)]{
      \includegraphics[width=0.154\linewidth]{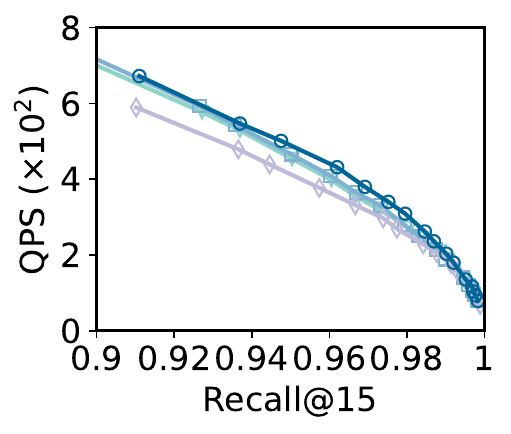}
     \label{fig:main_deep}
   }
   \hfill
   \subfigure[Crawl ($k=15$)]{
      \includegraphics[width=0.154\linewidth]{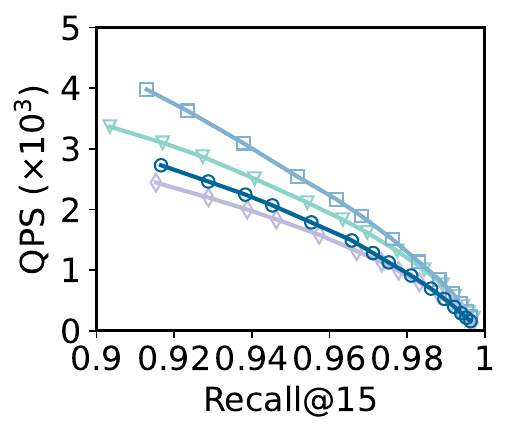}
     \label{fig:main_deep}
   }
   \hfill
   \subfigure[Sift1M ($k=50$)]{
      \includegraphics[width=0.154\linewidth]{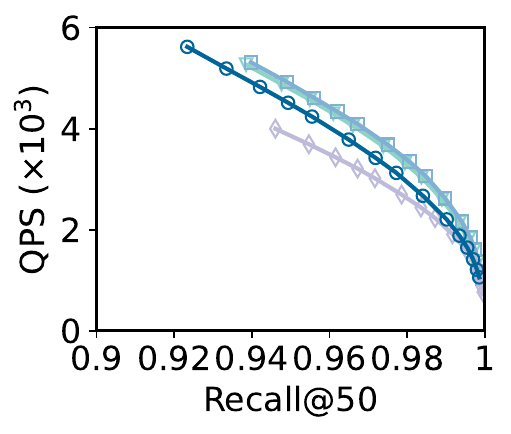}
     \label{fig:main_deep}
   }
   \hfill
   \subfigure[Gist1M ($k=50$)]{
      \includegraphics[width=0.154\linewidth]{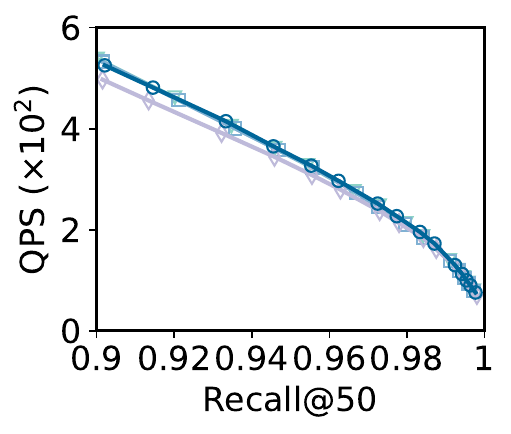}
     \label{fig:main_deep}
   }
   \hfill
   \subfigure[Crawl ($k=50$)]{
      \includegraphics[width=0.154\linewidth]{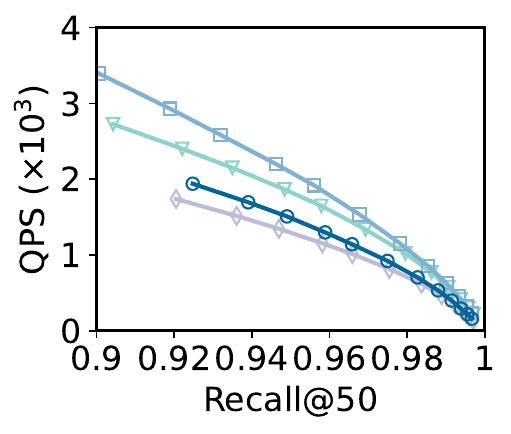}
     \label{fig:main_deep}
   }
  \vspace{-3mm}
  \caption{Comparisons in search performance on NSG and HNSW (Exp.2)}
  \vspace{-3mm}
  \label{fig:main}
\end{figure*}

\begin{figure*}[t]
\includegraphics[width=0.6\linewidth]{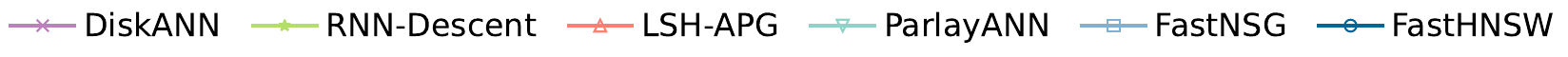}

  \centering
  \subfigure[Sift1M]{
    \includegraphics[width=0.154\linewidth]{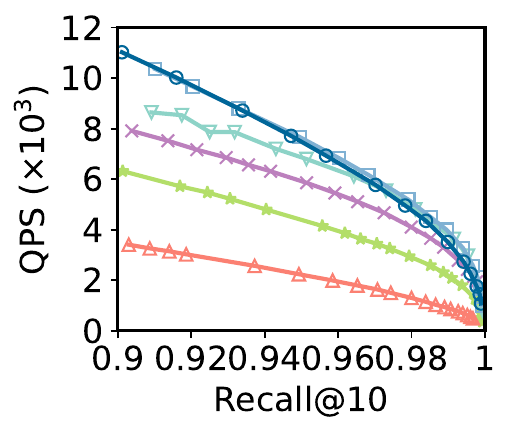}
    \label{fig:main_sift}
  }
  \hfill
  \subfigure[Gist1M]{
     \includegraphics[width=0.154\linewidth]{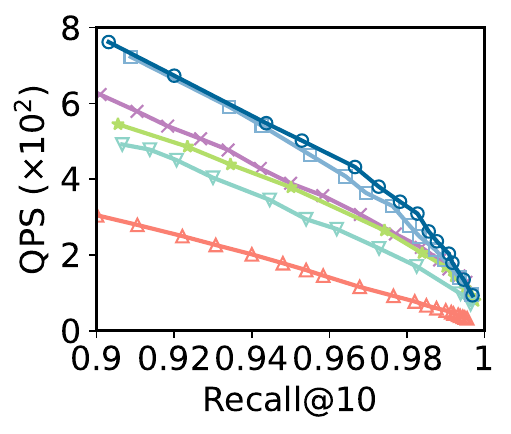}
    \label{fig:rmain_gist}
  }
  \hfill
  \subfigure[Msong]{
    \includegraphics[width=0.154\linewidth]{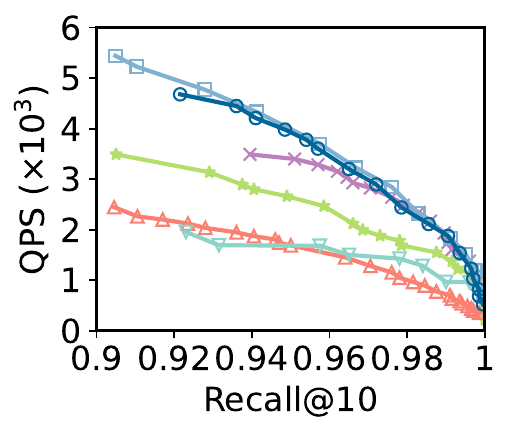}
    \label{fig:main_msong}
  }
  \hfill
  \subfigure[Crawl]{
     \includegraphics[width=0.154\linewidth]{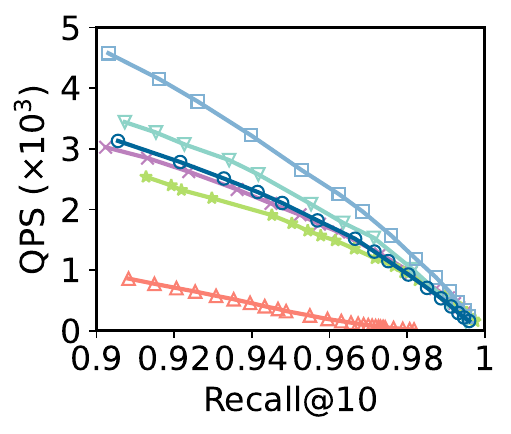}
    \label{fig:main_crawl}
  }
  \hfill
  \subfigure[Glove]{
     \includegraphics[width=0.154\linewidth]{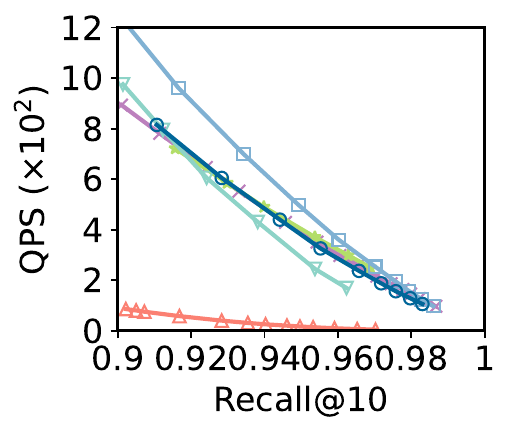}
    \label{fig:main_glove}
  }
  \hfill
  \subfigure[Deep1M]{
     \includegraphics[width=0.154\linewidth]{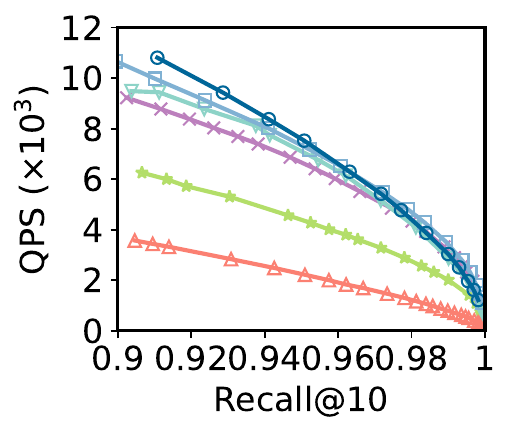}
    \label{fig:main_deep}
  }
  \vspace{-3mm}
  \caption{Comparisons in search performance between existing approaches and ours (Exp.4)}
  \label{fig:compare-existing}
\end{figure*}



\vspace{1mm}
\noindent \textbf{Exp.1: effects of the parameters.}
In the first experiment, we study the effects of the two newly proposed parameters in our approach: the number $I$ of iterations and the value of angle $\alpha$.

\textit{Exp.1-a: effects of $I$.}
In Algorithm \ref{alg:iter_nsg}, we iteratively conduct \texttt{K2P} and \texttt{P2K} before its termination.
We use \kw{FastNSG} as an example and show the result in Figure \ref{fig:iter_search}. We can see that more iterations improve the search performance but encounter marginal effects after the first two iterations.
Hence, the result shows $I$ affects the search performance of the finally derived proximity graph.
Since the building time increases as $I$ grows significantly as shown in Figure~\ref{fig:iter_build}, we set $I$ as $2$ by default in later experiments. 

\textit{Exp.1-b: effects of $\alpha$.} 
In this part, we study the effects of $\alpha$ on the $k$-CNA quality (measured by $recall$) and efficiency (measured by QPS). The results are shown in Figure~\ref{fig:alpha-exp}. Notably, when $\alpha = 60^{\circ}$, $\alpha$-pruning equals to the RNG pruning. We can see that as $\alpha$ increases slightly, $\alpha$-pruning enhances both the quality and efficiency of $k$-CNA compared with the RNG pruning. However, once $\alpha$ exceeds a specific value (e.g., $67^{\circ}$ on Glove), the efficiency of $k$-CNA degrades, due to more out-neighbors left by $\alpha$-pruning. Hence, $\alpha$ should be carefully tuned. In the following, we optimize $\alpha$ for each data via grid search, which starts from $60^{\circ}$.

\begin{figure}
    \centering
    \includegraphics[width=0.95\columnwidth]{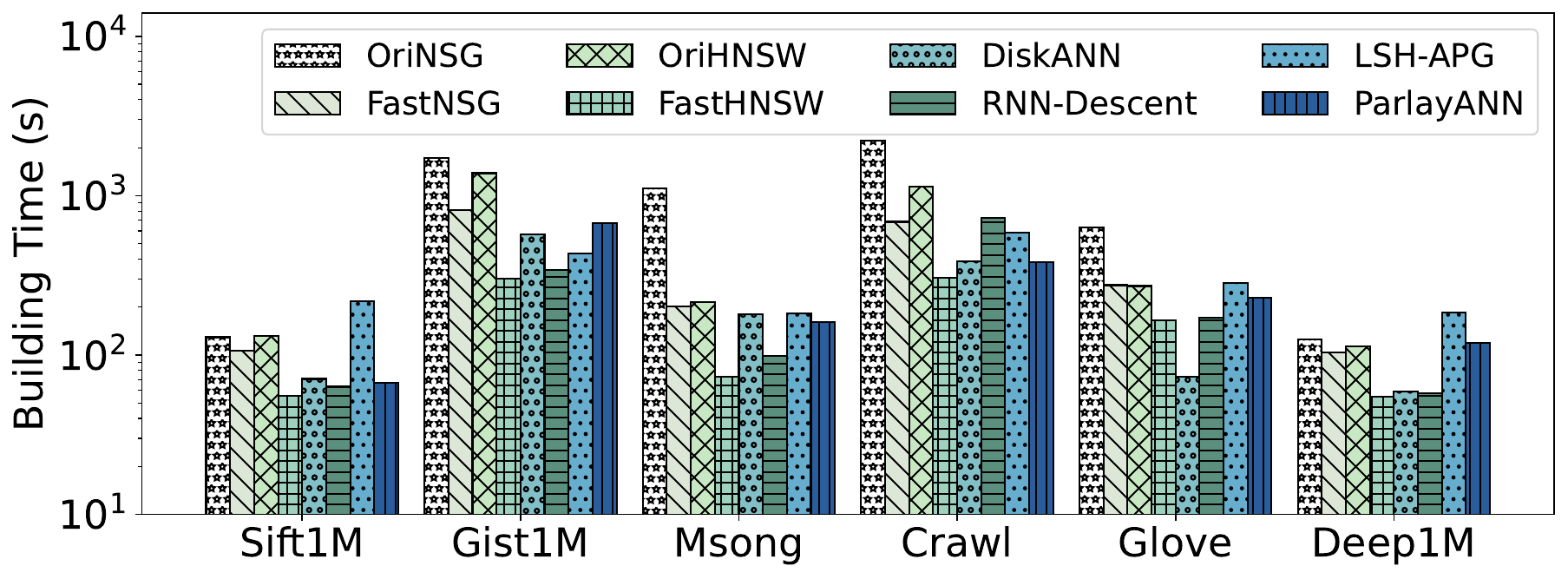}
    \vspace{-3mm}
        \caption{Comparisons in construction cost (Exp. 3 \& Exp. 4)}
        \vspace{-3mm}
  \label{fig:main_bc}
\end{figure}



\noindent \textbf{Exp.2 \& 3: main experimental results in search performance \& building time.}
In this part, we show the main results of this work by comparing our methods with the original algorithms in both search performance and building time.
We carefully choose the construction parameters for each method. For OriNsg, we adopt the construction parameters provided by the authors in \cite{Nsg} for Sift1M and Gist1M datasets. As for the Msong, Glove and Crawl datasets, we utilize the parameters provided by the recent survey \cite{survey2021}, which are determined using a grid search within the parameter space. We set the parameters of Deep1M as in \cite{diskann}.
For OriHNSW, we set the value of $M$ to be the same as in \cite{survey2021}, and the value of $ef$ is determined through a grid search within the parameter space.

The comparison of search performance and building time are presented in Figures \ref{fig:main} and \ref{fig:main_bc}, respectively. First, FastNSG exhibits comparable search performance to OriNSG in Figure \ref{fig:main}, while significantly reducing construction costs as shown in Figure \ref{fig:main_bc}. {Specifically, FastNSG achieves speedups of 1.2x, 2.1x, 5.6x, 3.2x, 2.3x and 1.2x over OriNSG on the Sift1M, Gist1M, Msong, Crawl, Glove and Deep1M datasets when we set $k=10$, respectively.} 
Second, FastHNSW shows significantly improved search performance compared to OriHNSW, {primarily due to obtaining more accurate $k$-CNA results. Note that FastHNSW finds candidates for each point on the entire dataset in each layer, while OriHNSW only on a subset (the existing nodes in the graph during insertions). } {Compared with OriHNSW, FastHNSW accelerates construction by 2.4x, 4.6x, 3.0x, 3.7x, 1.6x and 2.1x speedups on the six datasets when $k=10$.}

{Note that various approaches exhibit similar performance levels at high recall. This similarity arises from the fact that the search on proximity graph has two phases~\cite{vbase}, and the diverse structures of different proximity graphs mainly affect performance on the first phase due to variations of short-distance neighbors, while having slight impacts during the second phase, i.e., when recall is high.}

\begin{figure*}[t]
\includegraphics[width=0.4\linewidth]{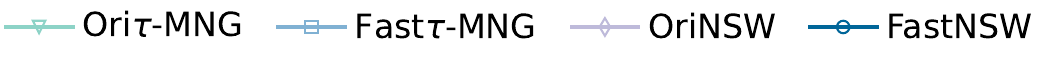}

  \centering
  \subfigure[Sift1M]{
    \includegraphics[width=0.154\linewidth]{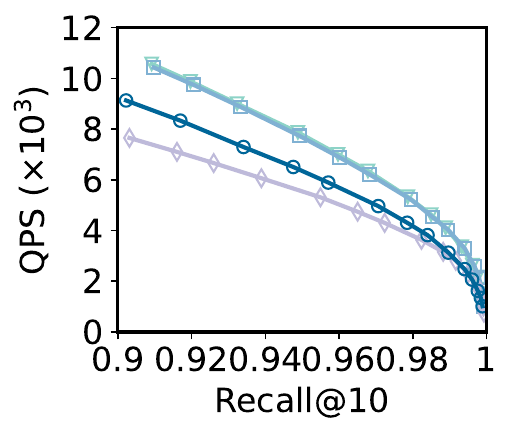}
    \label{fig:main_sift}
  }
   \hfill
   \subfigure[Gist1M]{
      \includegraphics[width=0.154\linewidth]{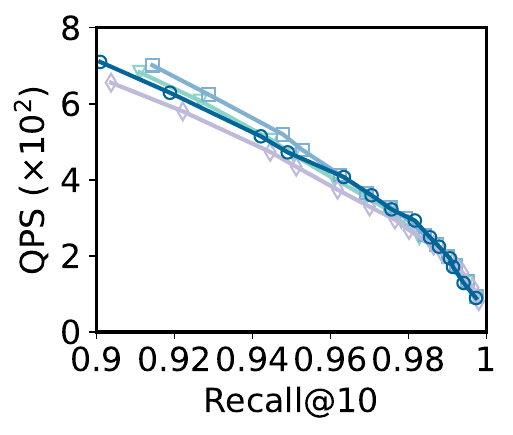}
     \label{fig:rmain_gist}
   }
  \hfill
  \subfigure[Msong]{
    \includegraphics[width=0.154\linewidth]{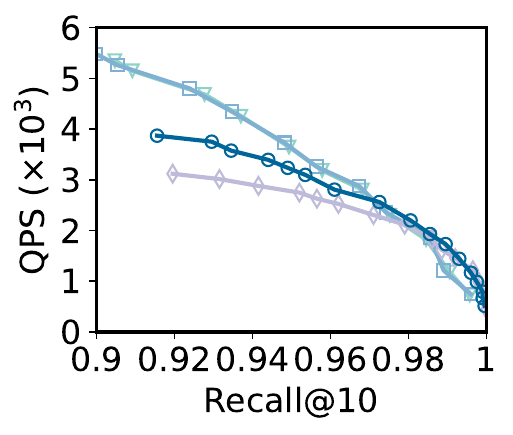}
    \label{fig:main_msong}
  }
  \hfill
  \subfigure[Crawl]{
     \includegraphics[width=0.154\linewidth]{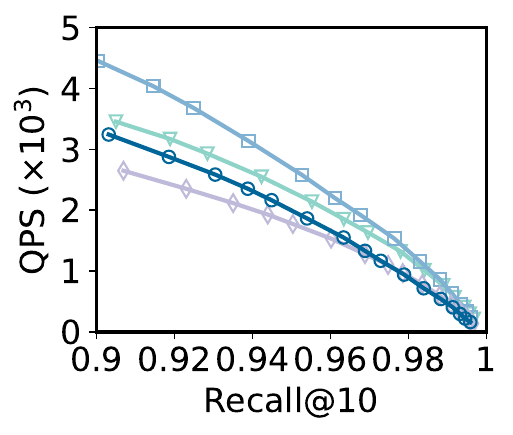}
    \label{fig:main_crawl}
  }
   \hfill
   \subfigure[Glove]{
      \includegraphics[width=0.154\linewidth]{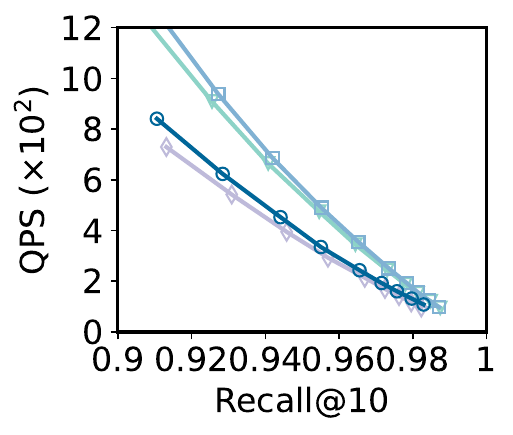}
     \label{fig:main_glove}
   }
   \hfill
   \subfigure[Deep1M]{
      \includegraphics[width=0.154\linewidth]{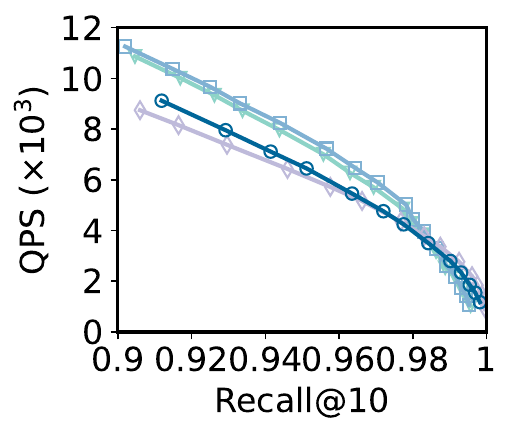}
     \label{fig:main_deep}
   }   
  \vspace{-3mm}
  \caption{Comparisons of other SOTA PGs (Exp.5)}
  \vspace{-3mm}
  \label{fig:extension}
\end{figure*}

\stitle{Exp.4: comparisons between existing approaches with ours.}
In this part, we compare our methods FastNSG and FastHNSW with other four recently proposed SOTA PG methods which focus on the index construction, i.e., DiskANN \cite{diskann}, LSH-APG \cite{Lsh-apg}, RNN-Descent \cite{OnoM23} and ParlayANN \cite{ManoharSBD0S024}. We show the comparisons of search performance in Figure \ref{fig:compare-existing} and that of building cost in Figure \ref{fig:main_bc}.
{Overall, the results show that our methods achieve much less construction cost while obtaining better search performance.}

\begin{figure}[t]
      \centering
      \begin{minipage}{0.48\linewidth}
        \includegraphics[width=0.95\columnwidth]{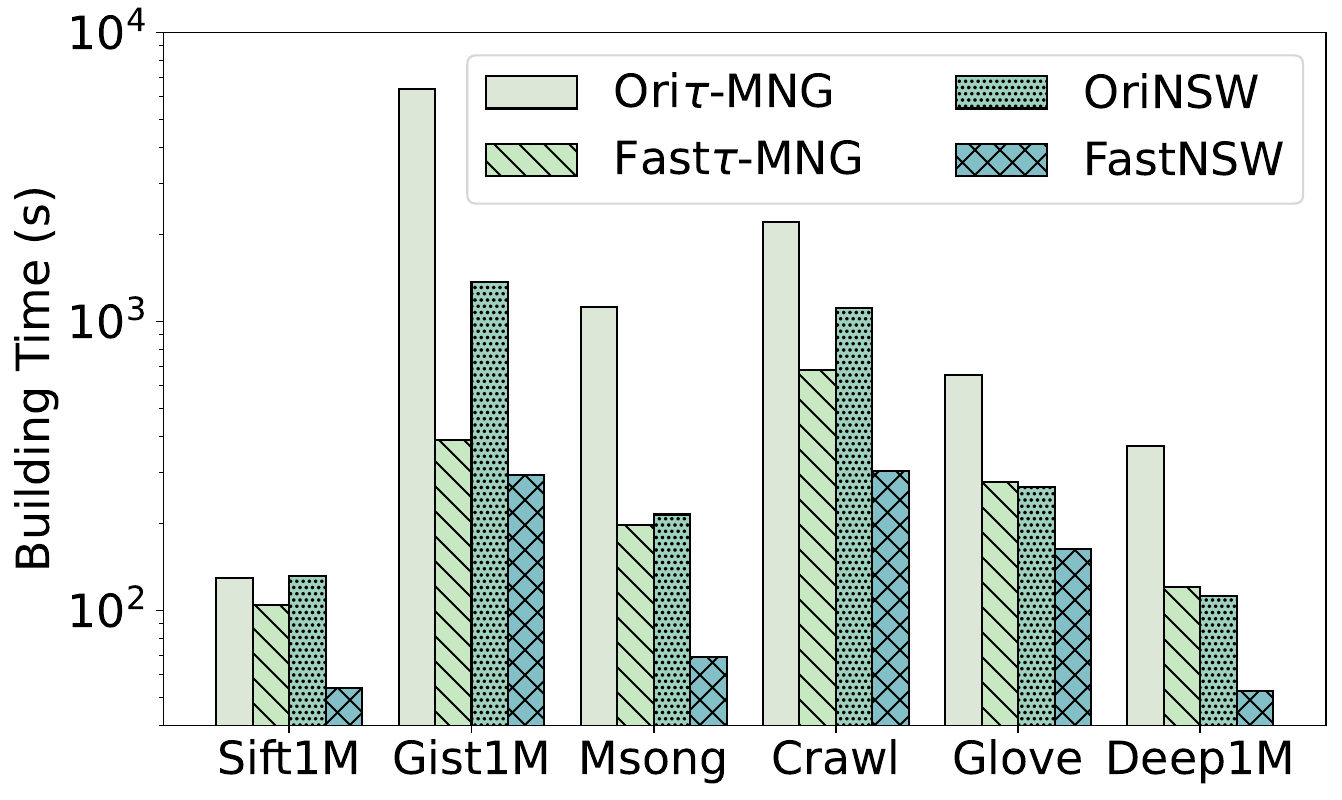}
    \caption{Construction time of other SOTA PGs (Exp.5)}
  \label{fig:extension_bc}
      \end{minipage}
      \hfill
      \begin{minipage}{0.48\linewidth}
    \includegraphics[width=0.98\columnwidth]{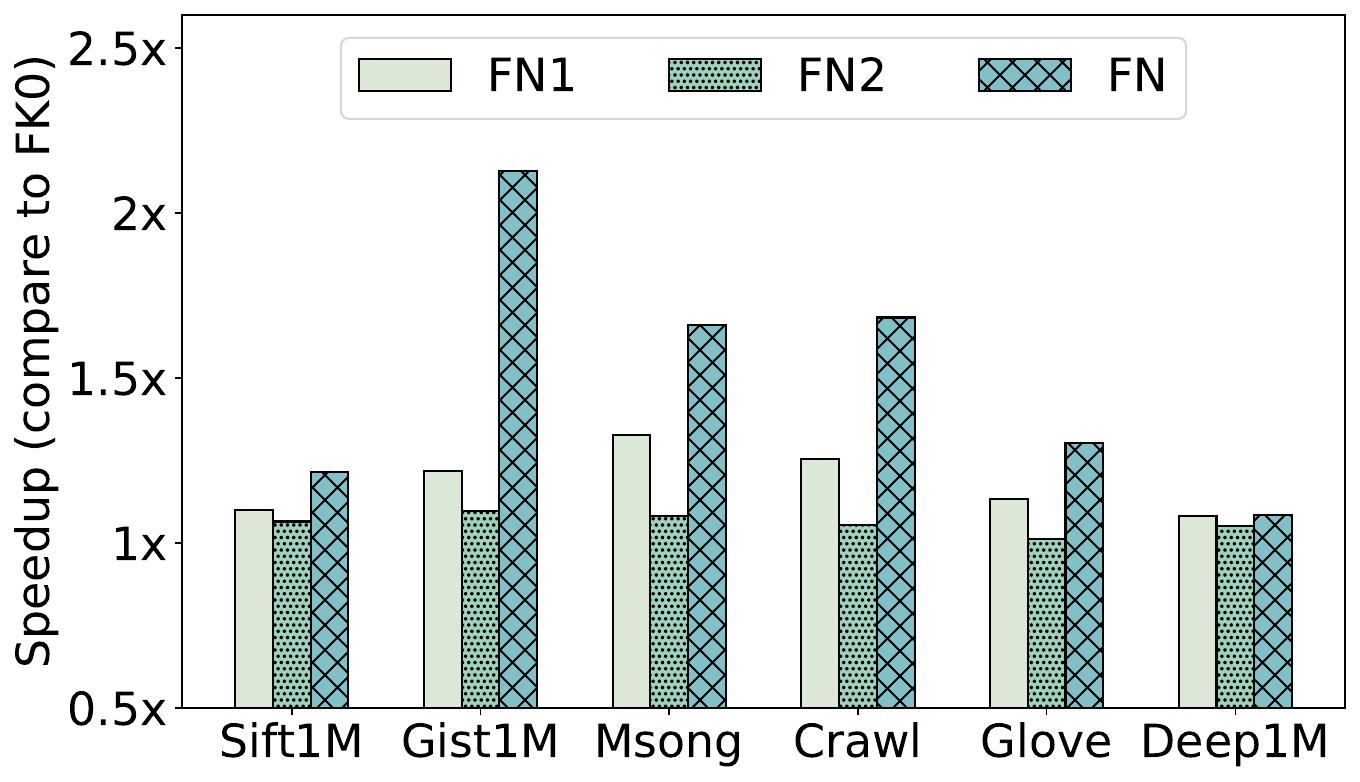}
    \caption{Effects of \kw{P2K} and \kw{K2P} techniques (Exp.6)}
  \label{fig:opt}
      \end{minipage}
      \vspace{-4mm}
\end{figure}


\stitle{Exp.5: extension of our framework on other SOTA PG approaches.} { In this part, we apply our framework to other two SOTA PG methods, i.e., $\tau$-MNG (another SOTA RNG method) and NSW (another SOTA NSWG method). As shown in Figure \ref{fig:extension}, our method Fast$\tau$-MNG achieves comparable or even better search performance compared with Ori$\tau$-MNG, while FastNSW significantly achieves better search performance than OriNSW. Moreover, as in Figure \ref{fig:extension_bc}, Fast$\tau$-MNG achieves construction speedups of 1.2x, 16.4x, 5.7x, 3.3x, 2.3x, 3.1x and 2.1x over Ori$\tau$-MNG on the six datasets respectively, while FastNSW obtains speedups of 4.6x, 3,1x, 3,6x, 1,6x, 2,1x and 2.1x respectively. Overall, our framework could be successfully applied to other SOTA PG methods, with superior construction efficiency and comparable search performance.}


{\noindent \textbf{Exp.6: effects of \kw{P2K} and \kw{K2P} techniques.}
{Here, we delve into the effects of our \kw{P2K} and \kw{K2P} techniques for \kw{FastNSG}. These techniques aim to reduce the redundant distance calculations of \kw{P2K} in consecutive iterations (referred to as \kw{Opt1}) and to reduce the redundant distance and angle computations of \kw{K2P} in consecutive iterations (referred to as \kw{Opt2}). These two strategies do not alter the final graph index, hence, we solely focus on the construction cost.
We evaluate four approaches: FN0, which lacks any optimization; FN1, integrating solely \kw{Opt1}; FN2, integrating solely \texttt{Opt2}; and FN, utilizing both two optimizations. The results are illustrated in Figure \ref{fig:opt}, where the time cost of each method is represented as the speedup over FN0.
Overall, each optimization greatly speeds up \kw{FastNSG}. These optimizations are independent of each other, with FN exhibiting the lowest construction cost. Besides, FN shows increasingly significant speedups as data dimensions increase, as both optimizations efficiently reduce the repeated computations.}}



\begin{figure}[t]
      \centering
      \begin{minipage}{0.46\linewidth}
        \includegraphics[width=0.95\columnwidth]{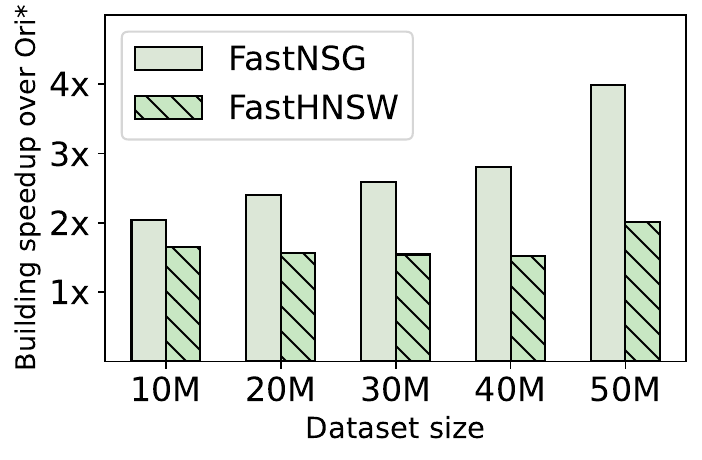}
    \caption{Scalability study of building index (Exp.7)}
  \label{fig:scal}
      \end{minipage}
      \hfill
      \begin{minipage}{0.50\linewidth}
      \includegraphics[width=0.95\columnwidth]{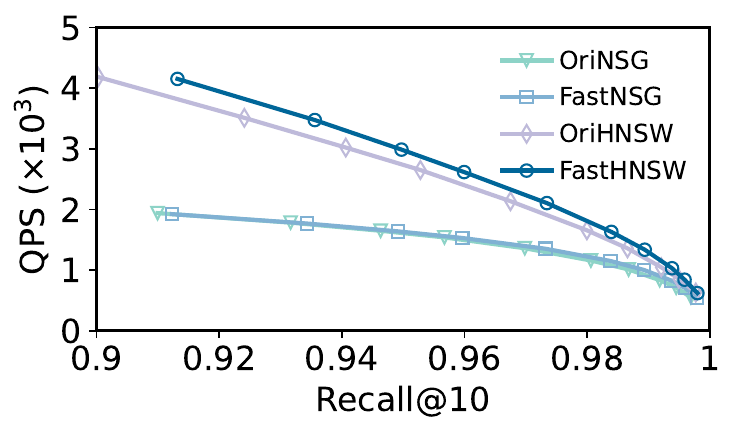}
    \caption{Scalability study of search on Sift50M (Exp.7)}
  \label{fig:scal50m}
      \end{minipage}
      \vspace{-4mm}
\end{figure}

{\noindent \textbf{Exp.7: scalability of our proposed methods.}
In this part, we assess the scalability of our methods on both index building and search performance using large data from Sift50M. As depicted in Figures \ref{fig:scal} and \ref{fig:scal50m}, our Fast* approaches consistently accelerate over the original Ori* methods as dataset sizes grow without compromising the search performance. Notably, the speedup in FastNSG further amplifies with expanding dataset sizes, underscoring the exceptional scalability of our approach. This demonstrates our RNG construction framework scales well as the data size rises. 
We put the search performance across four additional scales of the Sift50M datasets in the full version on our GitHub repository.
}

\section{Related Works}
\label{sec:rew}
There have been a bulk of works on processing $k$-ANN queries on high-dimensional data in the literature. To answer a $k$-ANN query, index structures are widely used to carefully select a small part of high-quality candidates and then verify them via distance computations, in order to return accurate results with little cost. According to recent experimental studies \cite{AumullerBF20,LiZAH20,Dpg,survey2021}, proximity graphs \cite{Nsg, Hnsw, Nssg} outperform other index structures such as hashing-based methods \cite{Datar2004, MPlsh, SKLSH, SRS}, inverted index-based methods \cite{PQ, IMI} and tree-based methods \cite{Kdtree,Flann} in search performance. Due to their excellent search performance, the SOTA PG methods such as NSG \cite{Nsg} and HNSW \cite{Hnsw} have been taken as the solution by industrial vector databases such as Milvus~\cite{milvus} and VBase~\cite{vbase}.



Moreover, several works that combine the parallel power of GPU and the filtering capacity of PG methods have been proposed~\cite{Song,Ggnn,Ganns,Cagra,liu2024retrievalattentionacceleratinglongcontextllm}. SONG \cite{Song} modifies the search method of existing methods such as HNSW in order to achieve higher throughput. Other GPU-accelerated methods such as GGNN \cite{Ggnn}, GANNS \cite{Ganns} and CAGRA \cite{Cagra} build their own graph index and have the corresponding search method.
In addition, there exist I/O-efficient PG-based approaches, such as DiskANN \cite{diskann} and Starling \cite{Starling}. Besides, some work expands $k$-ANN search to various scenarios, such as hybrid search \cite{vbase, GollapudiKSKBRL23, ZuoQZLD24}, out-of-distribution queries~\cite{abs-2211-12850}, and search on billion-scale datasets~\cite{ChenW21,NEURIPS2019_09853c7f,abs-2310-00402}.

\section{Conclusion}
\label{sec:clu}

{
In this paper, we study the efficient construction of the PG-based approaches for $k$-ANN search. We first analyze the importance of $k$-CNA quality for PG index and identify their issues on index construction by revisiting existing PG construction approaches.
To address these issues, we propose a novel construction framework for RNG with $\alpha$-pruning strategy and a self-iterative framework.
We then combine the layer-by-layer insertion strategy with our RNG construction framework to address the construction issue of HNSW (i.e., the representative NSWG).
Extensive experiments are conducted on real-world datasets to show the superiority of our methods. The results show our approaches exhibit a construction speedup to $5.6$x faster than the original methods of RNG and NSWG while delivering comparable or even superior search performance.
{
For future work, it is a promising direction to integrate a more efficient KNNG construction method into our initialization phase for further enhancing construction efficiency.
}

}

\bibliographystyle{ACM-Reference-Format}

\balance
\bibliography{main}

\clearpage
\appendix
\twocolumn[
\centerline{\Huge \bfseries Appendix} 
\vspace{\baselineskip} 
]

\section*{Additional Experiments}

\begin{figure}[h]
    \includegraphics[width=0.7\linewidth]{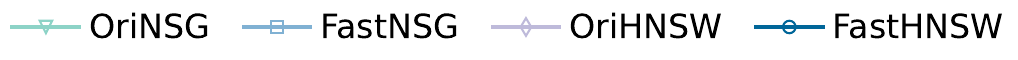}

  \centering
   \subfigure[Sift10M]{
    \includegraphics[width=0.47\linewidth]{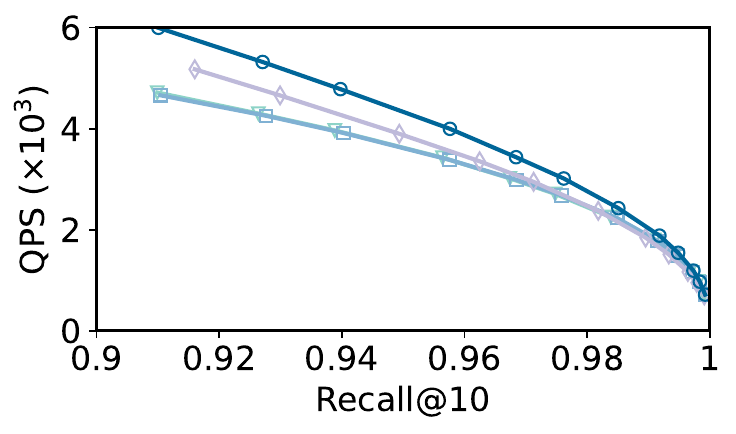}
  }
   \subfigure[Sift20M]{
    \includegraphics[width=0.47\linewidth]{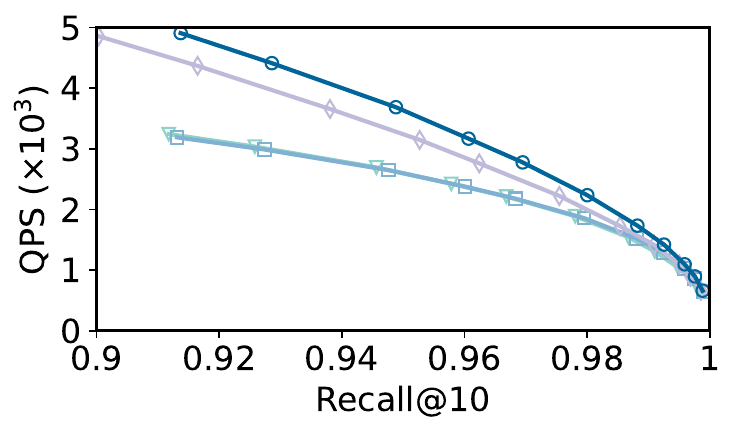}
  }
   \subfigure[Sift30M]{
    \includegraphics[width=0.47\linewidth]{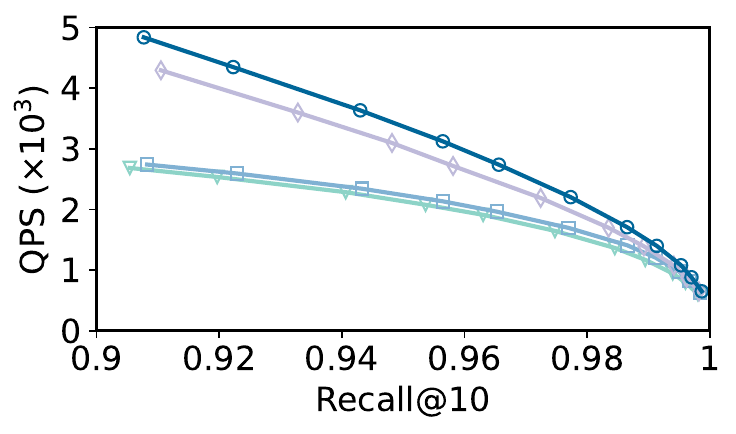}
  }
   \subfigure[Sift40M]{
    \includegraphics[width=0.47\linewidth]{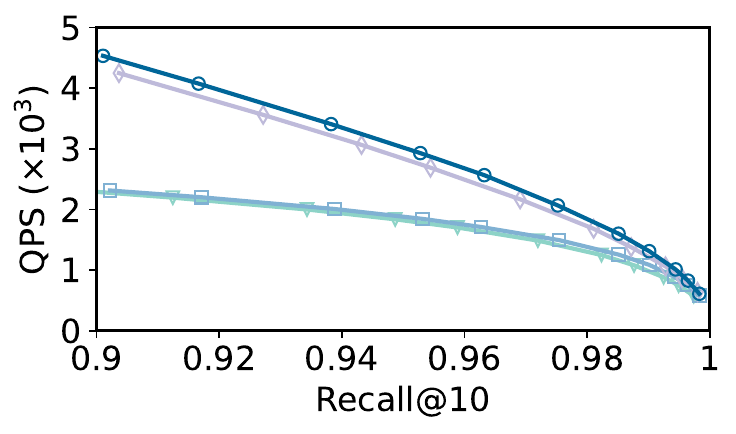}
  }
  \caption{Scalability study of search performance across four additional scales of Sift50M dataset (Exp. 7)}
\end{figure}

\begin{figure}[h]
      \centering
      \begin{minipage}{0.47\linewidth}
    \includegraphics[width=0.95\columnwidth]{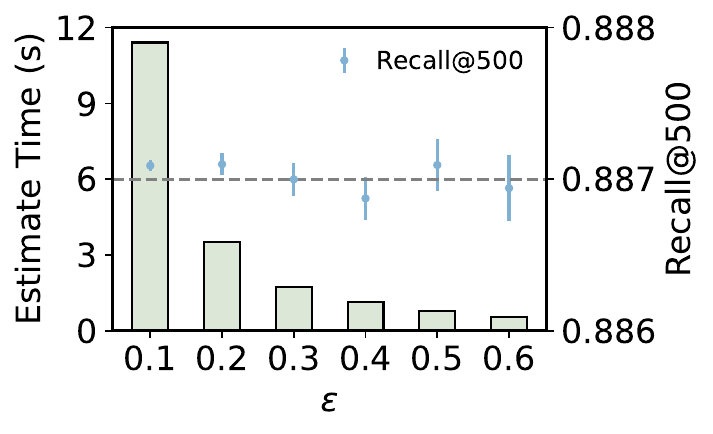}
    \caption{Recall estimation by random sampling (Exp.8)}
  \label{fig:random-recall}
      \end{minipage}
      \hfill
      \begin{minipage}{0.50\linewidth}
        \includegraphics[width=0.95\columnwidth]{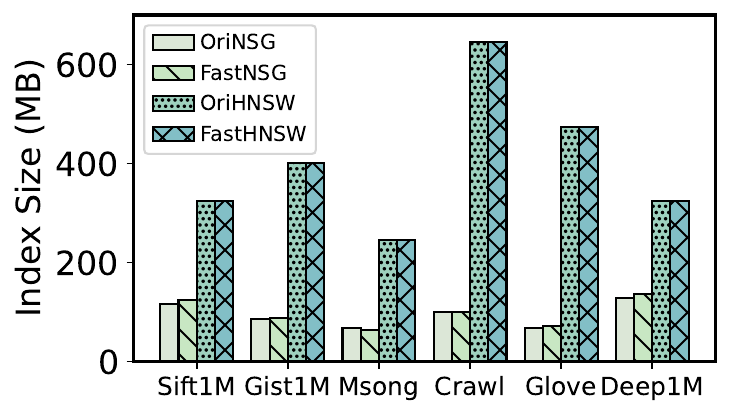}
    \caption{Index size comparisons (Exp.9)}
  \label{fig:index-size}
      \end{minipage}
\end{figure}

\vspace{1mm}
\noindent \textbf{Exp.8: KNNG recall estimation via random sampling.}
In this experiment, we study the impact of the parameter $\varepsilon$ on KNNG recall estimation (refer to Theorem~\ref{theo:eps}). The results on the Sift1M dataset are shown in Figure~\ref{fig:random-recall}, where the dotted line represents the exact value of recall@500 (the average recall of each node in KNNG ).
We find a reduction in running time as the value of $\varepsilon$ increases. While the fluctuation range (error) of recall estimation expands with higher $\varepsilon$ values, setting $\varepsilon$ to $0.6$ provides a reasonable estimate of recall swiftly, enabling a quick evaluation of KNNG quality. 

{
\vspace{1mm}
\noindent \textbf{Exp. 9: index size comparisons.}
We report the index sizes of our approaches compared to the original method of NSG and HNSW on all datasets, as depicted in Figure~\ref{fig:index-size}. The results indicate that our index closely resembles the original one, albeit slightly larger across all datasets.

\vspace{1mm}
\noindent \textbf{Exp. 10: other strategies for the node selection of upper layers in HNSW.}
In previous HNSW comparisons, we follow the original HNSW method for the node selection of upper layers by randomly determining the nodes in each upper layer, which is denoted as FastHNSW+Random here.
In this part, we explore the potential for a more refined approach to the node selection of each upper layer. However, given that the time required for such node selection is part of the index construction time, overly complex algorithms become impractical.

We introduce another strategy here, denoted as FastHNSW+Degree, which determines the nodes of upper layers in a bottom-up manner and iteratively selects nodes with maximum total out-degrees to the others in the current layer as the nodes of the next upper layer. We compare these two node selection strategies on Crawl and Deep1M datasets, with results detailed in Figure~\ref{fig:node_layer}.
FastHNSW+Degree enhances search performance on Crawl dataset but demonstrates inferior performance on Deep1M dataset. Additionally, FastHNSW+Degree demands more construction time, e.g., requiring an extra 2.8\% and 8.0\% building time on Crawl and Deep1M datasets respectively.

These results suggest the potential efficacy of alternative smarter strategies for node selection of upper layers in HNSW. However, it is vital to weigh such improvements against the added costs incurred during index construction. This highlights the need for further research to explore smarter node selection strategies for HNSW, and we leave this challenge as an open problem for future research.




\begin{figure}[h]
  \centering
    \includegraphics[width=0.7\linewidth]{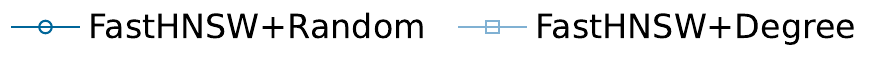}
  
   \subfigure[Crawl]{
    \includegraphics[width=0.47\linewidth]{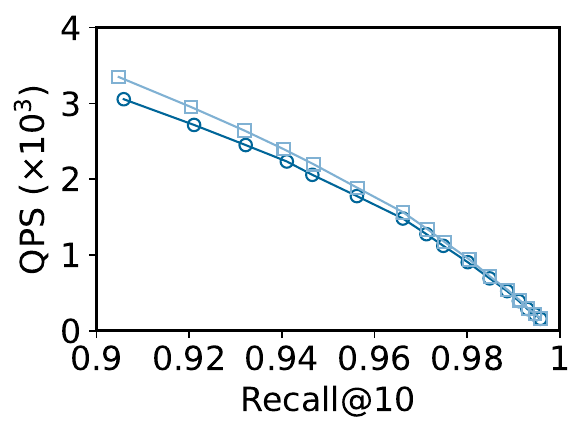}
  }
   \subfigure[Deep1M]{
    \includegraphics[width=0.47\linewidth]{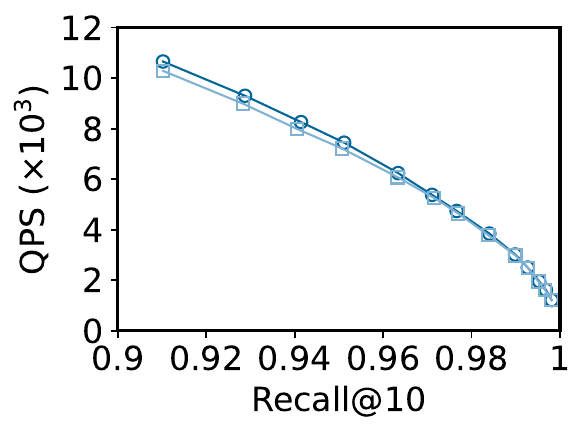}
  }
  \caption{Other strategies for the node selection of upper layers in HNSW (Exp. 10)}
  \label{fig:node_layer}
\end{figure}


\vspace{1mm}
\noindent \textbf{Exp. 11:  incremental evaluation on HNSW by replacing layer by layer.}
In order to incrementally evaluate our layer placement strategies on HNSW, where each layer in HNSW is replaced by RNG, we replace each layer in HNSW with RNG for experimental study. Specifically, the HNSWs constructed on Sift1M and Gist1M datasets consist of 4 layers, ranging from layer 0 to layer 3, with layer 0 being the lowest layer and containing nodes of the complete datasets. 
As depicted in Figure~\ref{fig:hnsw0layer}, we begin by replacing layer 3 in HNSW, labeled as Ori012+Fast3, followed by the replacement of both layers 2 and 3, denoted as Ori01+Fast23, then we replace layers 1,2 and 3 in HNSW, denoted as Ori0+Fast123, and finally all layers in original HNSW are replaced, denoted as FastHNSW.

The results indicate that replacing higher layers has only a minimal impact on search performance, whereas replacing the lowest layer, layer 0, significantly enhances search performance. This finding aligns with our analysis of NSWG construction issue, where the presence of long-distance edges in the lower layers due to incremental insertion is unnecessary in HNSW.

\newpage
\begin{figure}[h]
  \centering
\includegraphics[width=\linewidth]{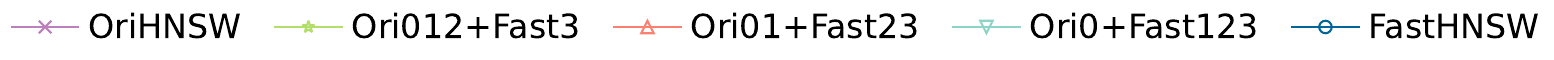}
  
   \subfigure[Sift1M]{
    \includegraphics[width=0.47\linewidth]{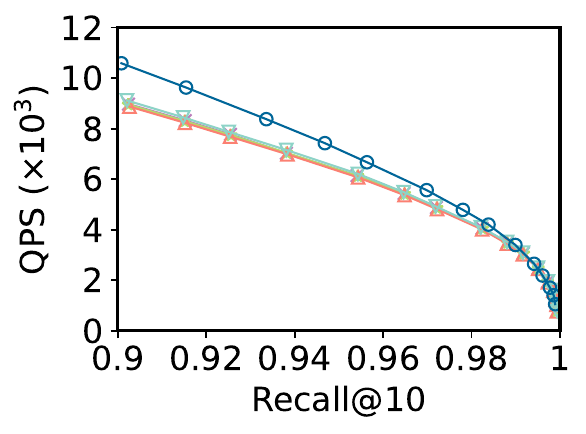}
    \label{fig:swaplayer_sift}
  }
   \subfigure[Gist1M]{
    \includegraphics[width=0.47\linewidth]{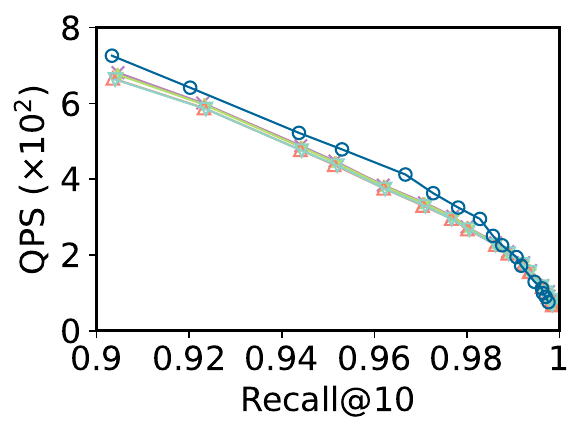}
    \label{fig:swaplayer_gist}
  }
  \caption{ The search performance of the graph replacing layers of OriHNSW with that of FastHNSW (Exp. 11)}
  \label{fig:hnsw0layer}
\end{figure}

\vspace{1mm}
\noindent \textbf{Exp.12: effects of the initial KNNG quality.}
To explore the effects of the initial KNNG quality on the final $k$-CNA results, we conduct additional experiments as follows. In these experiments, during the initialization phase, we change the number $iter$ of iterations of KGraph~\cite{KGraph} to vary both its construction cost and quality. Specifically, when $iter=0$, during initialization, $k$ data points are randomly selected as neighbors for each data point. Next, when increasing the value of $iter$, at each iteration, every data point computed the distance between itself and its 2-hop neighbors to identify new, closer $k$ data points as neighbors.
Following this, our framework was employed to iteratively enhance the KNNG and conduct searches on the refined KNNG (i.e., PG) to generate a new KNNG.

The results are depicted in Figure~\ref{fig:kgiter}, where the initial point of each line represents the time and quality of the KNNG derived from the initialization phase with different iterations. Each subsequent point along the line denotes the time and quality of the obtained KNNG at each iteration within our framework. The results illustrate that when striving for high-quality $k$-CNA results, e.g., the recall@500 is close to 1, the time remains steady across various settings of $iter$ for small $iter$ values, e.g., $iter\le 8$. However, as $iter$ increases, the time required may significantly escalate compared to instances where $iter$ is small, e.g., $iter=14$ in Gist1M.

Therefore, based on the above experimental analysis, it appears that constructing a high-quality KNNG in the initialization phase is not a compulsory step within our framework to achieve high-quality $k$-CNA results. However, there exists the potential to develop a more efficient method for constructing a high-quality KNNG that outperforms KGraph, the current state-of-the-art KNNG construction approach. Integrating such an approach into the initialization phase of our framework could further enhance the construction efficiency. This possibility is left open as a potential direction for future research.

\begin{figure}[h]
  \centering
   \subfigure[Gist1M]{
    \includegraphics[width=0.47\linewidth]{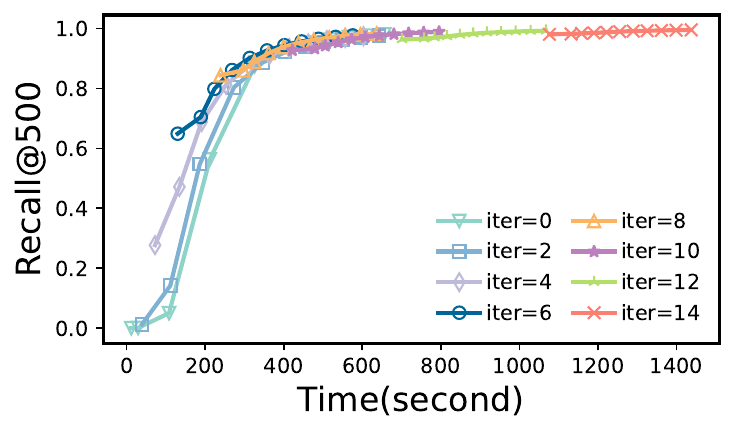}
    \label{fig:kgiter_gist}
  }
   \subfigure[Crawl]{
    \includegraphics[width=0.47\linewidth]{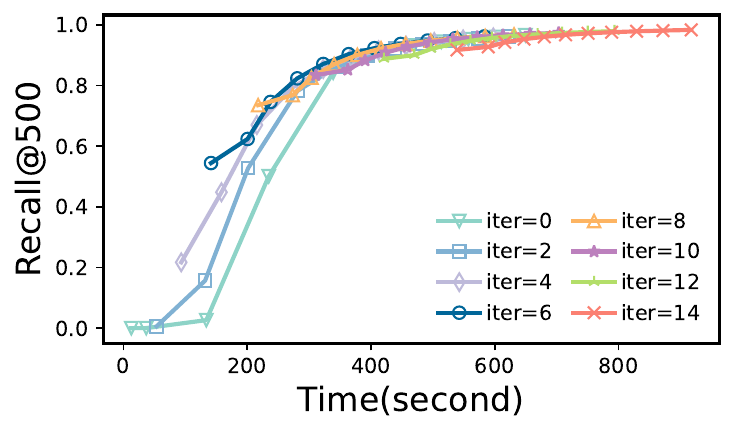}
    \label{fig:kgiter_crawl}
  }
  \caption{The effects of KGraph construction cost on the quality of $k$-CNA results (Exp. 12)}
  \label{fig:kgiter}
\end{figure}
}

\end{document}

%% file: custom_sytles.tex
\usepackage{graphicx}
\usepackage{balance}  
\usepackage{lipsum}
\usepackage{amsmath}
\usepackage{physics}
\usepackage{subfigure}
\usepackage{tabularx}
\usepackage{color}
\usepackage{colortbl}
\usepackage{float}
\usepackage{listings}
\usepackage{multirow}
\usepackage[normalem]{ulem}
 \usepackage{longtable}
\usepackage{enumitem}
\usepackage{multirow}
\usepackage{booktabs}
\usepackage{capt-of}
\usepackage{pifont}
\usepackage{ulem}
\usepackage{soul}
\usepackage{cancel}
\usepackage{bm}
\usepackage{url}
\usepackage{caption}

\def\BibTeX{{\rm B\kern-.05em{\sc i\kern-.025em b}\kern-.08em
    T\kern-.1667em\lower.7ex\hbox{E}\kern-.125emX}}

 \graphicspath{{./Graph/}, {./Fig/}, {./Legend/}}

\long\def\comment#1{}

\newcounter{example}[section]
\renewcommand{\theexample}{\nthesection.\arabic{example}}
\newenvironment{example}{
     \refstepcounter{example}
     {\vspace{1ex} \noindent\bf  Example  \theexample:}}{
     \vspace{1ex}} 

\newcounter{definition}[section]
\renewcommand{\thedefinition}{\nthesection.\arabic{definition}}

\newcounter{theorem}[section]
\renewcommand{\thetheorem}{\nthesection.\arabic{theorem}}
\newenvironment{theorem}{\begin{em}
        \refstepcounter{theorem}
        {\vspace{1ex} \noindent\bf  Theorem  \thetheorem:}}{
        \end{em}\vspace{1ex}} 

\newcounter{lemma}[section]
\renewcommand{\thelemma}{\nthesection.\arabic{lemma}}
\newenvironment{lemma}{\begin{em}
        \refstepcounter{lemma}
        {\vspace{1ex}\noindent\bf Lemma \thelemma:}}{
        \end{em}\vspace{1ex}} 

\newcounter{corollary}[section]
\renewcommand{\thecorollary}{\nthesection.\arabic{corollary}}

\newcounter{proposition}[section]
\renewcommand{\theproposition}{\nthesection.\arabic{proposition}}

\newcounter{remark}[section]
\renewcommand{\theremark}{\nthesection.\arabic{remark}}

\newcommand{\proofsketch}{\noindent{\bf Proof Sketch: }}

\newcommand{\nthesection}{\arabic{section}}

\newcommand{\eop}{\hspace*{\fill}\mbox{$\Box$}\vspace*{1ex}}

\newcommand{\stitle}[1]{\noindent{\bf #1}}

\newcommand{\green}[1]{\textcolor{green}{}}

\newcommand{\kw}[1]{{\ensuremath {\mathsf{#1}}}\xspace}

\newcommand\bigutimes{\mathop{\ooalign{$\bigcup$\cr%
   \hfil\raise0.36ex\hbox{$\scriptscriptstyle\boldsymbol{\times}$}\hfil\cr}}}
\newcommand\bigumius{\mathop{\ooalign{$\bigcup$\cr%
   \hfil\raise0.36ex\hbox{$\scriptscriptstyle\boldsymbol{-}$}\hfil\cr}}}

